\documentclass[aps,twocolumn,showpacs]{revtex4}
\usepackage{graphicx}
\usepackage{textcomp}
\usepackage{psfrag}
\begin{document}
\title{Influence of Visco-elastic Nature on the Intermittent Peel Front Dynamics of the Adhesive Tape}
\author{Jagadish Kumar}
\author{G. Ananthakrishna}
\affiliation{Materials Research Centre, Indian Institute of Science, Bangalore 560012, India}

\begin{abstract}
We investigate the influence of visco-elastic nature of the adhesive on the intermittent peel front dynamics by extending a recently introduced model for peeling of an adhesive tape. As time and rate dependent deformation of the adhesives are measured in stationary conditions,  a crucial step in  incorporating the visco-elastic effects applicable to unstable intermittent peel dynamics is the introduction of a dynamization scheme that eliminates the explicit time dependence in terms of dynamical variables. We find contrasting influences of visco-elastic contribution in different regions of tape mass, roller inertia, and pull velocity. As the model acoustic energy dissipated depends on the nature of the peel front and its dynamical evolution, the combined effect of the roller inertia and pull velocity makes the acoustic energy noisier for small tape mass and low pull velocity while it is burst-like for low tape mass, intermediate values of the roller inertia and high pull velocity. The changes are quantified by calculating the largest Lyapunov exponent and analyzing the statistical distributions of the amplitudes and durations of the model acoustic energy signals. Both single and two stage power law distributions are observed. Scaling relations between the exponents are derived which show that the exponents corresponding to large values of event sizes and durations are completely determined by those for small values. The scaling relations are found to be satisfied in all cases studied. Interestingly, we find only five types of model acoustic emission signals among multitude of possibilities of the peel front configurations.
\end{abstract}

\maketitle

\section{Introduction}
\label{sec1}
Science of adhesion is truly interdisciplinary involving  a great variety of different interrelated physical phenomena such as  visco-elastic, visco-plastic deformation, mechanics of contact,  fracture and interfacial properties such as debonding and rupture  of adhesive bonds. Detailed mechanisms that control various properties of such a complicated mixture of phenomena are not yet well   understood. Substantial part of our understanding of adhesion is based on near equilibrium or stationary state experiments supplemented by the corresponding theoretical analysis. However,  we routinely encounter situations that represent time dependent or dynamical manifestations of adhesion such as  use  of adhesive tapes for packing and sealing. Current understanding of dynamical aspects of adhesion is largely based on a few types of experiments such as peeling of adhesive tapes under constant pull velocity or constant load conditions \cite{MB,BC97,CGVB04}. In both cases, the peel process is intermittent accompanied by a characteristics audible noise \cite{MB,BC97,CGVB04}. Further, observations of the peel front under controlled conditions also reveal that the peel front exhibits fibrills \cite{McEwan,Urahama,Dickinson}. These experiments show that the intermittent peel process results from switching of the peel process between the two stable branches that are  separated by an unstable branch not accessible to experiments. The low velocity branch is attributed to visco-elastic dissipation while that at high velocities  to the crack speed reaching the Rayleigh wave velocity. However, it must be emphasized that these two branches are measured in stationary state situations.

Our earlier attempts to understand the peel front dynamics \cite{Rumi06,Jag08a,Jag08b} were  focused on understanding the origin of the intermittent peeling of an adhesive tape and its connection to acoustic emission (AE). The basic idea was to describe the peel front dynamics by writing down an appropriate Lagrangian that includes contributions from the kinetic energy, the potential energy, and the dissipation arising from rapid movement of the peel front apart from that subsumed in the bistable peel force function. The latter depends on the local displacement rate of the peel front, and is a crucial input for describing the spatio-temporal peel front instability \cite{Rumi06,Jag08a,Jag08b}. Indeed, the basic premise of our model  is that acoustic emission is the energy dissipated during abrupt stick-slip events given by the spatial average of the square of the gradient of displacement rate\cite{Rumi06,Jag08a,Jag08b}. 

We demonstrated that the model was able to predict  a number of experimentally observed features  \cite{Rumi06,Jag08a,Jag08b}. For instance, the stuck-peeled (SP) configuration in the model \cite{Rumi06,Jag08a,Jag08b} mimics the fibrillar pattern of the peel front observed in experiments \cite{McEwan,Dickinson,Urahama}. Further, several statistical and dynamical features of acoustic emission such as the transition from burst to continuous type signal observed in experiments was also predicted by the model. In addition, the two stage power law distribution for the amplitudes of the experimental AE signals along with the associated exponent values were also reproduced by the model. The model also predicts spatio-temporal chaos for a specific set of parameters \cite{Rumi06,Jag08a,Jag08b}. This also suggests that AE signals may have a hidden signature of chaotic dynamics, which has also been verified \cite{Jag08a,Jag08b}.  Indeed, crucial insight  into  many  observed experimental features of acoustic emission has been provided by the model {\it that  establishes a correspondence  between the nature of the peel front and acoustic energy dissipated.} This correspondence helps us to understand the mechanisms that control the crossover from burst type to continuous type AE signals observed with increasing pull velocity \cite{CGVB04,Jag08a,Jag08b}.

Apart from the peel problem, number of stick-slip systems such as sliding friction \cite{Heslot94,Persson}, the Portevin-Le Chatelier (PLC) effect \cite{PLC,GA07}, nonlinear rheological response of  micelles \cite{Sood06}  display negative 'force-velocity' relationship.  Except in the case of sliding friction \cite{Heslot94}, the existence of the unstable branch is only inferred. For instance in the case of the PLC effect, a kind of plastic instability observed during tensile deformation of dilute alloys  \cite{GA07,Anan04}, the measured strain rate sensitivity of the flow stress shows the two stable dissipative branches only \cite{Kubin86}.

Models that attempt to explain the dynamical features of stick-slip systems use the macroscopic phenomenological negative force-velocity relation as input although the unstable region is not accessible. This is true in the present case also \cite{Rumi06,Jag08a,Jag08b}. In general, the negative force - velocity relationship is attributed to rate dependent deformation. The underlying physical cause  is the strong history dependent nature of the deformation of these materials.  This property well documented in the case of  adhesives \cite{Kae64,GP69,TIY04}. Indeed, the two dissipative branches reflect precisely the rate dependence. However, from a dynamical point of view, stick-slip dynamics usually results from a competition between intrinsic time scales, possibly several (for example, inertial times scales of the tape mass and roller inertia in our model),  and the applied time scale  \cite{Persson,BKC, Anan06,GA07}. As the adhesive is a visco-elastic material, both the visco-elastic time scale and rate dependent deformation of the adhesive \cite{Kae64,GP69,TIY04} are expected to  play an important role in the peel dynamics.

However, the time and rate dependence features have not been included in our recent model for the peel front dynamics \cite{Rumi06, Jag08a, Jag08b}.  Moreover, our earlier studies show that different regions of the peel front experience different peel velocities and thus the rate dependent deformation of adhesives \cite{Kae64,GP69,TIY04} can be important.  Further, the peel process is itself sensitive to the interplay of various time scales in the model and thus, the nature of peel front dynamics  will be influenced by the additional visco-elastic time scale.  Thus, our primary aim is to examine the role played by the visco-elastic nature of the adhesive on the peel front dynamics and its influence on  acoustic emission. However, as conventional rate dependent effects are always measured in stationary conditions,  and our secondary aim, although a prerequisite,  is to devise a suitable  framework for including time and rate dependent effects  in unstable intermittent dynamical situations. Clearly, such an approach should be useful in understanding the rate dependent effects in the other bistable force-velocity situations. 

\section{The Model}
\label{sec2}

\begin{figure}
\hbox{
\centering
\includegraphics[height=2.5cm,width=4.8cm]{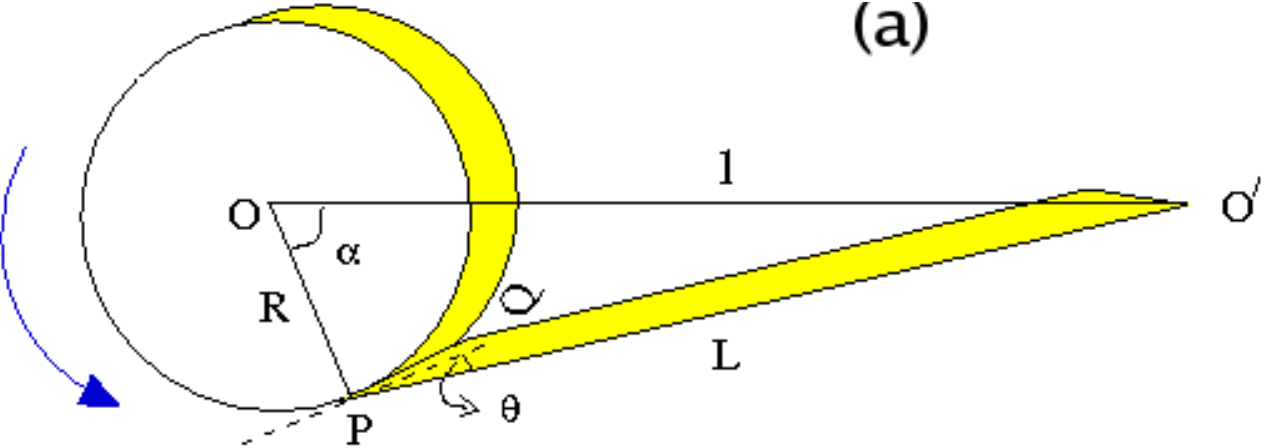}
\includegraphics[height=3.0cm,width=3.5cm]{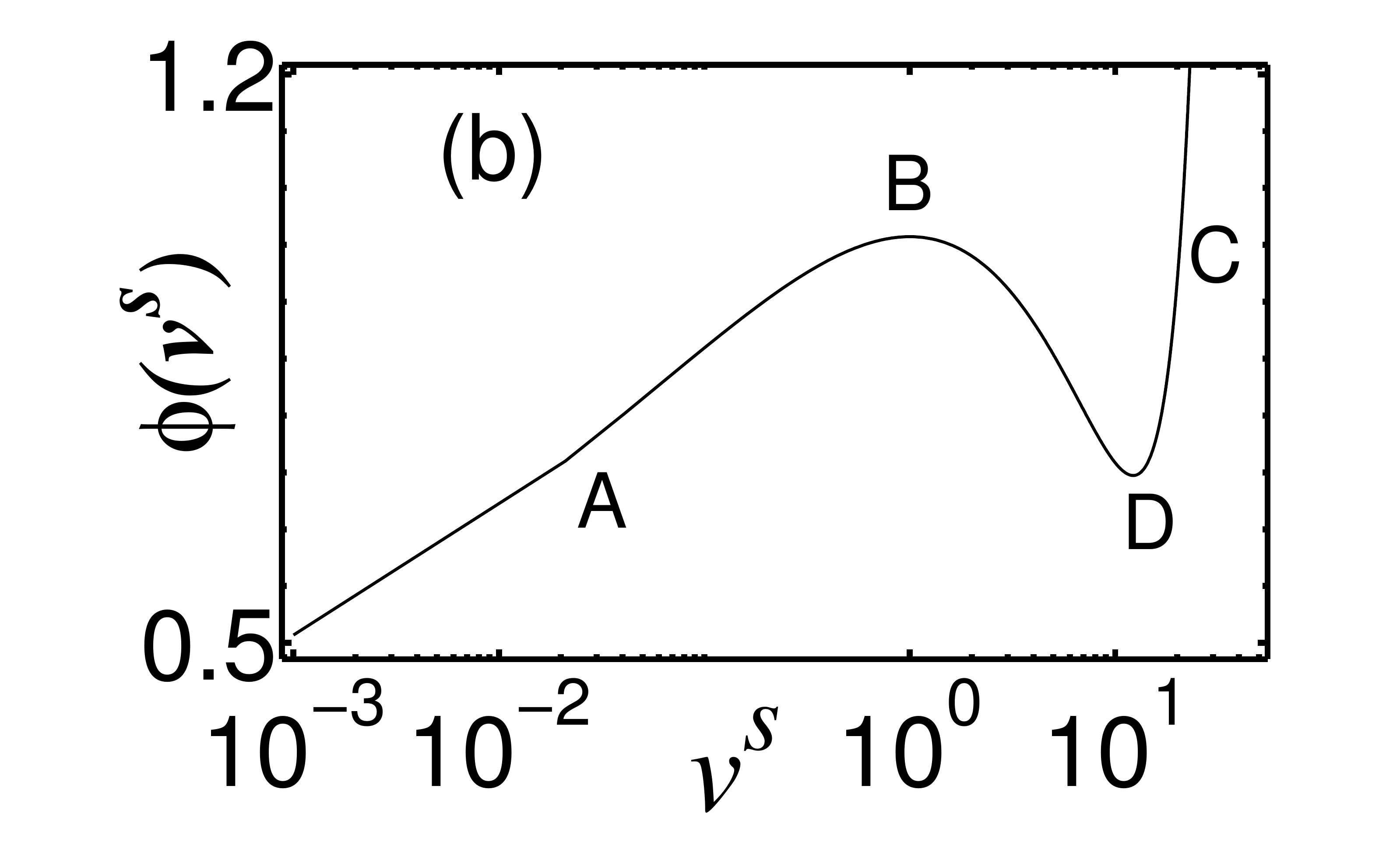}
}
\caption{(Color online a) (a) A schematic representation of the experimental setup. (b) Plot of the scaled peel force function $\phi(v^s)$ as a function of scaled peel velocity $v^s$.
}
\label{tapewidth}
\end{figure}

In our model, the visco-elastic nature of the adhesive was included indirectly by assuming  an effective spring constant for the peel front that was taken to be constant $k_0$, small compared to the spring constant of the tape material $k_t$.  In reality, the soft adhesive should be described by a time dependent spring constant that is conventionally described by assuming a single relaxation time scale given by  
\begin{equation}
k_g(t) = k_g(0) + [k_g(\infty) -k_g(0)] exp( -t/T_a),
\label{kg}
\end{equation}
where $k_g(0)$ is the spring constant for short times while $k_g(\infty) (>> k_g(0))$ is the saturation value  and $T_a$ is the visco-elastic time scale. (A more general expression that includes several relaxations times can also be written down in terms of network models  \cite{YZZ04}.) 
Note that this equation has proper limits, namely, for short times, $k_g(t) \rightarrow k_g(0)$ while for long times $k_g(t) \rightarrow k_g(\infty)$. The time variable, however,  is measured from some reference state that implies that Eq. (\ref{kg}) saturates quickly in a time interval long compared to the relaxation time $T_a$. On the other hand, our  equations of motion  support {\it oscillatory solutions} that capture the intermittent stick-slip dynamics of the peel front. Hence, {\it the explicit dependence on time in Eq. (\ref{kg}) that makes no reference to slow-fast intermittent dynamics would saturate within a few cycles of stick-slip.} Thus, we need to devise a method to include the visco-elastic effects valid for the intermittent state. To accomplish this, we eliminate the time variable in terms dynamical variables as the model equations are autonomous.

While Eq. (\ref{kg}) is often taken to represent rate dependent deformation of adhesives, there is no rate dependence in its present form. The latter is a complex phenomenon \cite{Kae64,GP69,TIY04},  commonly observed in visco-elastic and visco-plastic materials, as also in plastic deformation of metals and alloys \cite{GA07}.  The imposed deformation rate limits the internal relaxation processes which themselves are functions of local strain, local strain rate, local stress, temperature, deformation mode etc \cite{GA07,Kae64,GP69,TIY04} \footnote{Most constitutive forms of rate dependent deformation are phenomenological. They are suited for the limited purpose they are designed. One simple way to include rate dependence valid in stable deformation conditions is to assume that $T_a$ depends on the applied rate.}.  
To the best of our knowledge, no theoretical approach has been developed for including  rate dependent effects in unstable intermittent flow situations.  {\it Here we propose an algorithm that is suitable for this situation.} 

What is required is a method of including the following features of rate dependence  of the adhesive. For instance, when a certain segment of the peel front experiences low velocities,  the segment should undergo visco-elastic creep as there is enough time for the adhesive to relax.  In contrast, when a segment experiences high peel velocity, the adhesive segment should behave like a solid as there is very little time for the visco-elastic relaxation to occur. This physical implication of rate dependent deformation of the adhesive is however not captured by Eq. (\ref{kg}) since time variable enters explicitly while the model equations display intermittent slow-fast dynamics. Here, we propose a method of incorporating the rate dependence by eliminating time in terms of the two relevant dynamical variables, namely, velocity and displacement. Rewriting  Eq. ({\ref{kg}}) in terms of $v$ and $u$, we have
\begin{eqnarray}
 k_g(u/v) &=& k_g(0) + [k_g(\infty)-k_g(0)]exp (-\frac{u}{v T_a}).
\label{glue0}
\end{eqnarray}
Clearly, Eq. (\ref{glue0}) captures the desired rate dependent deformation of visco-elastic peel front as both the local peel velocity $v$ and displacement $u$ depend on the imposed pull velocity. We refer to this equation as {\it dynamized form } of Eq. (\ref{kg}). ( The sense in which  dynamization is used here is very different from that used earlier \cite{Rumi04} or in Ref. \cite{Kubin}. In the latter, an explicit dependence on applied strain rate is introduced into the negative strain rate sensitivity of the flow stress. A similar approach is adopted in Ref. \cite{Rumi04}.) 
If we choose $k_g(0)$ to be small compared to $k_g(\infty)$, it is easy to check that  when a peel segment experiences low velocity  (in the region of the left branch of the peel force function), the behavior of the adhesive is viscous liquid like (i.e., $k_g \sim k_g(0)$).  It is solid like when the peel segment is on the high velocity branch (i.e., $k_g \sim k_g(\infty)$). Thus, the spring constant is made dynamical. 

We begin by collecting some relevant geometrical details. Experiments carried out under constant pull velocity have a set up similar to the schematic  shown in Fig. $1(a)$.   An adhesive roller tape of radius $R$ mounted on an axis passing through $O$ is pulled at a constant velocity $V$ using a  motor positioned at $O'$ at a distance $l$ from  $O$. Let the peeled length of the tape $PO'$ be $L$. From the figure it is clear that the tangent to the contact point $P$ (representing the contact line $PQ$) subtends an angle $\theta$ to the line $PO'$ and the line $PO$ subtends an angle $\alpha$  with the horizontal $OO'$ at $O$. Then geometry of the set up leads to  $L\ {cos}\, \theta = -l\ {sin}\,\alpha$ and 
$L\ {sin}\,\theta = l\ { cos}\,\alpha - R$. As the peel point $P$ moves with a local velocity $v$, the pull velocity has to satisfy the relation
\begin{equation}
V=v + \dot u + R\dot \alpha \  {\rm cos}\ \theta \ ,
\label{Localconstr} 
\end{equation}
where $u$ is the displacement. Let $u(y)$ be the displacement with respect to the uniform  `stuck' peel front along the peel front direction. Similarly, we define all the relevant variables  $v(y),\theta(y),\alpha(y)$  at every point $y$ along the contact line. Then, as the entire tape of width $b$ is pulled at the velocity $V$, a more general equation holds
\begin{eqnarray}
{1\over b} \int^b_0 \big[V- v(y) -  \dot u(y)  -  R\dot\alpha(y) \ \ {\rm cos} \ \ \theta(y) \ \ \big]dy =0. 
\label{Vconstraint}
\end{eqnarray}

The model is described by a Lagrangian that has contributions from the kinetic energy, the potential energy  and frictional dissipative terms. The total kinetic energy of the system $U_k$ is the sum of the rotational kinetic energy of the roller tape and the kinetic energy of the stretched part of the  tape. This is given by
\begin{equation}
U_K={1\over2}\int^b_0 \xi \big[\dot \alpha(y) +{v(y)\over R} \big]^2 dy + {1\over2}\int^b_0 \rho \big[\dot u(y) \big]^2 dy.
\end{equation}
Here, $\xi$ is the moment of inertia per unit width of the roller tape and $\rho$ is the mass per unit width of the tape. The total potential energy $U_p$ consists of the contribution from the displacement of the peel front due to stretching of the peeled tape and possible inhomogeneous nature of the peel front. Thus,
\begin{equation}
U_P={1\over2}\int^b_0 {k_t\over b} \Big[u(y) \Big]^2 dy + {1\over2}\int^b_0 {k_g(t) b} \Big[{\partial u(y) \over \partial y} \Big]^2 dy. 
\end{equation}

The peel process always involves dissipation. The dominant contribution comes from the  peel force function that describes the two stable branches separated by an unstable one connecting the two. In addition, we include another dissipative mechanism  that is {\it crucial for describing acoustic emission as well as peel front instability.} This term  arises from  the accelerated motion of local regions of the peel front during the abrupt rupture process. We  consider   this term to be responsible for the generation of acoustic signals \cite{Rumi06}. Any rapid movement of the rupture front generates dissipative forces that tend to resist the motion of the slip. Such dissipative forces  are modeled by  
\begin{equation}
{\cal R}_{AE}= {1\over2}\int^b_0 {\Gamma_u\over b} \Big[{\partial \dot u(y) \over \partial y} \Big]^2 dy.
\label{RAE}
\end{equation}
Note that  this term  has the same form as the acoustic wave energy   generated by dislocations during  plastic deformation. This is given by $E_{ae} \propto \dot \epsilon^2(r)$, where $\dot \epsilon(r)$ is  the local plastic strain rate \cite{Rumiepl}. Therefore,  we interpret  ${\cal R}_{AE}$ as the energy dissipated in the form of acoustic emission. Indeed, such a dissipative term has proved useful in explaining the power law statistics of the AE signals during martensitic transformation \cite{vives,Rajeev01,Kala03} as also in explaining certain AE features in fracture studies of rock sample \cite{Rumiepl} apart from the AE features in the peel problem \cite{Rumi06,Jag08a,Jag08b}. Then, the total dissipation is  
\begin{equation}
{\cal R}={1\over b} \int^b_0 \int f(v(y)) dv dy + {\cal R}_{AE},
\label{diss}
\end{equation}
where $f(v)$ physically represents the peel force function assumed to be derivable from a potential function  $\Phi(v) = \int f(v)dv$ (see Ref. \cite{Rumi05}). The form of the peel force function is represented by  
\begin{equation}
f(v)=402v^{0.34}+171v^{0.16}+68e^{(v/7.7)}-369.65v^{0.5}-2.
\label{frictioncurve}
\end{equation}
The above form preserves major experimental features such as the magnitude of the velocity jumps across the two branches of $f(v)$, the range of  values of the measured peel force function and its value at the onset of the unstable branch.

We now write the equations in a non-dimensional form.  We define  a time like variable $\tau = \omega_{u} t$ with $\omega_{u}^2={k_t/(b \ \rho)}$ and  a length scale  $d=f_{max}/k_t$, where  $f_{max}$ is the value of $f(v)$ at $v_{max}$ on the left stable branch. Using these variables, we define scaled variables $u = X d = X (f_{max}/k_t) $, $l =  l^s d$, $L =  L^s d$   and $R =  R^s d$. The peel force $f(v)$ can be written as  $\phi(v^s) = f(v^s(v))/f_{max}$, where the dimensionless peel and pull velocities are given  by  $v^s=v/v_c\omega_u d$ and $V^s=V/v_c\omega_u d$ respectively. Here, $v_c = v_{max}/ \omega_u d$ represents the dimensionless critical  velocity at which the unstable branch starts. Using this we can define  a few relevant  scaled parameters $C_f=(f_{max}/k_t)^2(\rho/\xi)$,   $k=k_g(u/v) b^2/(k_t a^2)$,  $\Delta k=(k_g(\infty) -k_g(0)) b^2/(k_t a^2)$,  $\gamma_u = \Gamma_u \omega_u/(k_t a^2)$, and $y = ar$, where $a$ is a  unit length variable along the peel front.

Then, the scaled local form of  Eq. (\ref{Localconstr}) takes the form 
\begin{equation}
\dot X = (V^s - v^s)v_c + R^s \ {l^s \over L^s} \ ({sin}\ \alpha)\ \dot \alpha. 
\label{localconstraint}
\end{equation}  
In terms of the scaled variables, the scaled kinetic energy $U^s_K$ and scaled potential energy $U^s_P$ are respectively given by 
\begin{eqnarray}
U^s_K &= &{1\over2 C_f}\int^{b/a}_0 \Big[ \Big(\dot \alpha(r) +{v_c v^s(r)\over R^s} \Big)^2  + C_f{\dot X}^2(r)\Big] dr, \\
\label{ScKE}
U^s_P &=& {1\over2}\int^{b/a}_0 \Big[ X^2(r) + k(X/v^s) \Big({\partial X(r) \over \partial r} \Big)^2 \Big]dr. 
\label{ScPE}
\end{eqnarray} 
The total dissipation in the scaled form is
\begin{equation}
{\cal R}^s=R_{f}^s+R_{AE} = {1\over b} \int^{b/a}_0 \Big[\int \phi(v^s(r)) dv^s  + {\gamma_u\over2}\Big({\partial \dot X(r) \over \partial r} \Big)^2 \Big]dr.
\label{ScDiss}
\end{equation}
The first term on the right-hand side is the frictional dissipation arising from the peel force function. The second term is the scaled form of the acoustic energy dissipated. The scaled peel force function, $\phi(v^s)$, can be obtained by using the scaled velocities in Eq. (\ref{frictioncurve}). The nature of $\phi(v^s)$ is shown in Fig. \ref{tapewidth}(b). Note that the maximum occurs at $v^s =1$. We shall refer the left branch AB as the "stuck state" and  the high velocity branch CD as the "peeled state".

Finally, in scaled variables, Eq. (\ref{glue0}) takes the form
\begin{eqnarray}
k_g(X/v^s) &=& k_g(0) + [k_g(\infty)-k_g(0)]exp (-\frac{X}{v^s\tau_a}),
\label{glue}
\end{eqnarray}
where $\tau_a=\frac{v_{max}T_a}{d}$. The Lagrange equations of motion in terms of the generalized coordinates $\alpha(r), \dot\alpha(r), \dot X(r) $ and $\dot X(r)$ are

\begin{eqnarray}
\ddot \alpha &=& - {v_c \dot v^s \over R^s}  - C_f R^s {{l^s \over L^s} \, {sin}\, \alpha \over (1 + {l^s \over L^s} \, {sin}\, \alpha)} \phi(v^s),\label{seqalpha} \\
\nonumber
\ddot X &=& - X + k(X/v^s) {\partial^2 X \over \partial r^2}+ {\phi(v^s)\over (1+ {l^s\over L^s} \, {sin}\, \alpha)} + \gamma_u {\partial^2 \dot X \over \partial r^2}\\
\nonumber
&+&{\Delta k\over {2 \tau_a v^s}}\Big({\partial X \over \partial r}\Big)^2 e^{-{X\over {v^s\tau_a}}}-{\Delta k\over {2 \tau_a v_c}} \Big[{1\over {v^s}^2}\dot X \Big({\partial X \over \partial r}\Big)^2 \\
\nonumber
&& e^{-{X\over {v^s\tau_a}}}\Big(1-{X\over {v^s\tau_a}}\Big) +{{X\dot v^s}\over {v^s}^4} \Big({\partial X \over \partial r}\Big)^2 e^{-{X\over {v^s\tau_a}}}\\
&&\Big({X\over \tau_a}-2v^s \Big) +{2X\over {v^s}^2} \Big ({\partial X \over \partial r}\Big) \Big({\partial \dot X \over \partial r}\Big)e^{-{X\over {v^s\tau_a}}} \Big]. 
\label{sequ}
\end{eqnarray}
However,  Eqs. (\ref{seqalpha}) and (\ref{sequ}) should  also satisfy  the constraint Eq. (\ref{localconstraint}). A standard way of implementing the consistency condition between equations (\ref{seqalpha}), (\ref{sequ}) and (\ref{localconstraint}) is to use the theory of  mechanical  systems with constraints \cite{ECG}. This leads to an  equation for the acceleration  variable $\dot v^s (r)$ (obtained by differentiating Eq. (\ref{localconstraint}) and using Eq. (\ref{sequ})), given by
\begin{eqnarray}
\nonumber
\dot v^s v_c &=& \Big[{1\over{1-{\Delta k\over {2 \tau_a v_c^2 {v^s}^4}}  X \Big ({\partial X \over \partial r}\Big)^2 e^{-{X\over {v^s\tau_a}}}\Big({X\over \tau_a}-2v^s\Big)}}\Big]\\
\nonumber
&& \Big[{\Delta k\over {2 \tau_a v_c}} \Big({1\over {v^s}^2}\dot X \Big({\partial X \over \partial r}\Big)^2 e^{-{X\over {v^s\tau_a}}}(1-{X\over {v^s\tau_a}}) \Big)\\
\nonumber
&+&{\Delta k\over { \tau_a v_c}}\Big ({X\over {v^s}^2}\Big ) \Big ({\partial X \over \partial r}\Big) \Big({\partial \dot X \over \partial r}\Big)e^{-{X\over {v^s\tau_a}}}\\
\nonumber
&-&{\Delta k\over {2 \tau_a v^s}} \Big({\partial X \over \partial r}\Big)^2 e^{-{X\over {v^s\tau_a}}}+X-k(X/v^s){\partial^2 X \over \partial r^2}\\
\nonumber
&-& {\phi(v^s)\over (1+ {l^s\over L^s} \, {sin}\, \alpha)} -\gamma_u {\partial^2 \dot X \over \partial r^2} + {R^sl^s\over L^s} \Big(\dot \alpha^2 (cos \alpha \\
&-&{R^s l^s}({sin \alpha \over L^s})^2 ) +  sin \alpha {\ddot \alpha} \Big) \Big]. 
\label{sdotv1}
\end{eqnarray}
Henceforth, we will drop the denominator in the above equation as  ${\Delta k\over {2 \tau_a v_c^2 {v^s}^4}}  X \Big ({\partial X \over \partial r}\Big)^2 e^{-{X\over {v^s\tau_a}}}\Big({X\over \tau_a}-{2v^s\Big)} << 1$.

Equations (\ref{localconstraint}), (\ref{seqalpha}) and (\ref{sdotv1}) constitute a set of nonlinear partial differential equations that determine the dynamics of the peel front. They have been solved by discretizing the peel front on a grid of  N points using an adaptive step size stiff differential equations solver (MATLAB package).  We have used open boundary conditions appropriate for the problem. The initial conditions are drawn from the stuck configuration, i.e., the values are from the left branch of $\phi(v^s)$ with a small spatial inhomogeneity in $X$ such that  they satisfy Eq. (\ref{localconstraint}) approximately. The system is evolved till a steady state is reached before the data is accumulated. 

\section{Dynamics of the peel front}

\subsection{Competing time scales and parameters} 

The dynamics of the model is sensitive to the four time scales (in the reduced variables) determined by the parameters $C_f$, $\gamma_u$, $k$ and $V^s$. $C_f$ is related to the ratio of the inertial time of the tape mass to that of roller inertia. $k$ is the ratio of spring constant of the glue to that of the tape in a dynamical state and  $\tau_a$ is the visco-elastic time scale. The dissipation parameter $\gamma_u$ reflects the rate at which the local strain rate relaxes.  Finally, the pull velocity $V^s$ determines the duration over which all the internal relaxations are allowed to occur. The range of $C_f$ is determined by the allowed values of the tape mass $m$ and the roller inertia $I$. Following our earlier studies,  $I$ is varied from $10^{-5} $ to $10^{-2}$,  and $m$ from  0.001 to 0.1.  Thus, $C_f$ can be varied over a few orders of magnitude keeping one of them fixed. The range of $V^s$ of interest is determined by the instability domain.

In our numerical simulations we have taken $\tau_a =0.526$,  $k_g(0) \sim 0.01 k_t$ and $k_g(\infty) \sim 0.71 k_t$. These values fix $\Delta k=(k_g(\infty) -k_g(0)) b^2/(k_t a^2)$. A rough estimate of the variation of the elastic constant of the adhesive in a dynamical situation (i.e., during the time evolution of the equations) can be obtained by inserting typical values of the peel force function. For example, for low values of $v^s \sim 10^{-3}$, $k_g \sim 0.01 k_t$, while  for values of $\phi(v^s)$ at its maximum, namely $v^s=1$ and $X=1$, we get   $k_g(X/v^s) = 0.1146 k_t$. This value is of the same order as the value of $k_0=0.1 k_t$ used in our earlier studies ($k_t=1000N/m$) \cite{Jag08b,Jag08a,Rumi06}. On the right branch, taking $v^s=20$ and $X=1$, we get $k_g =0.647 k_t$. Thus, the variation of the spring constant in  dynamical situations can be substantial. Indeed, such a large variation of the modulus  of the adhesive material is known from rate dependent studies \cite{Kae64,GP69,TIY04}.

To understand the nature of acoustic emission, we begin with a few observations about the model acoustic energy $R_{AE}$. From  Eq. (\ref{ScDiss}), it is clear that  the acoustic energy $R_{AE}$ is the spatial average of the square of the gradient of the displacement rate.  However, the  peel front configurations are sensitive to the  value of $\gamma_u$.  From Eq. (\ref{sdotv1}), low $\gamma_u$ implies that the coupling between neighboring sites is weak  and hence the local dynamics dominates. Thus, the displacement rate at one spatial location has enough freedom to  deviate from that of its neighbor. This generally leads to stuck-peeled configurations. In contrast, as shown in Ref. \cite{Rumi06,Jag08b}, high $\gamma_u$ implies  strong near neighbor coupling and thus leads to smooth synchronous peel front, and consequently sharp bursts are seen in the model acoustic energy $R_{AE}$.

In view of this, we shall fix $\gamma_u$ at a low value. This choice is also supported by the estimate presented in  \cite{Jag08b}. The unscaled dissipation parameter $\Gamma_u$ is related to the fluid shear viscosity $\eta$ \cite{Land}. Using  typical values of $\eta$ for adhesives, it was shown that the order of magnitude estimate of $\gamma_u$ is $\sim 10^{-3} -10^{-4}$. The results presented here are for $\gamma_u =0.01$ as the peel front patterns  for smaller $\gamma_u$  are similar. Finally, we note that the exact nature of the peel front pattern and the associated model acoustic energy $R_{AE}$ depends on other parameters  $C_f$ and the pull velocity $V^s$.

Finally, we estimate the region of time scales where the visco-elastic time scale influences the dynamics. In the unscaled variables, we have two frequencies $\omega_u = (k_t/m)^{1/2}$ and $\omega_{\alpha} = (Rf/I)^{1/2}$.  Our earlier study has demonstrated that low mass limit of the ODE model \cite{Rumi05} corresponds to the differential algebraic equations \cite{Rumi04}. In this limit, we have shown that the orbit in the $X-v^s$ plane jumps abruptly across the two stable branches of the peel force function amounting to infinite acceleration \cite{Rumi05,Jag08b}.  On the other hand, finite tape mass causes jumps in $v^s$ to occur over a finite time scale. This often restricts the phase space trajectory from  visiting the high velocity branch of $\phi(v^s)$  \cite{Rumi06,Jag08a,Jag08b}. From this point of view, the visco-elastic time  scale should be expected to influence the dynamics at low and intermediate tape mass values. However, we stress that the roller inertia $I$ also influences the dynamics. 

\subsection{Methods of analysis} 

Simple dynamical tools such as velocity-space-time patterns, phase plots in the $X-v^s$ plane for an arbitrary spatial point on the peel front and the associated model acoustic energy $R_{AE}$ are quite useful in studying  the influence of visco-elasticity of the adhesive on the peel front dynamics.  Our earlier studies suggest that the system of equations could be spatio-temporally chaotic (STC) for a certain set of parameters values.  This  can be  quantified by calculating the largest Lyapunov exponent (LLE) from the equations of motion.  We will also use statistical tools such as calculating the distributions of amplitudes and durations of the fluctuating acoustic energy signals. As shown earlier, the distribution of event sizes often exhibit power law behavior\cite{Rumi06,Jag08a,Jag08b}. 

\subsubsection{Dynamical tools}

Our earlier work has established that the equations of motion for the $k_0$-model are spatio-temporally chaotic for a certain range of parameters. The largest Lyapunov exponent for such systems should be positive.  Thus, in principal, one expects to find a range of values of the parameters for which the visco-elastic model also to be spatio-temporally chaotic (which  however could be different from those of the $k_0$- model).   In the following we briefly describe the method of calculating the largest Lyapunov exponent. 

Lyapunov exponents are measures of the sensitivity to initial conditions.   Positive Lyapunov exponent is a measure of the rate of divergence of near-by orbits.  This is calculated by choosing two orbits that are close to each other and evolving them for a certain interval of time. If the system is chaotic, the orbits diverge from each other in a short time along the directions corresponding to positive Lyapunov exponents and contract along directions corresponding to negative Lyapunov exponents. Here, the phase space is $4N$ dimensional and there are as many Lyapunov exponents. For a spatio-temporal chaotic  system, the number of positive Lyapunov exponents scales with the system size \cite{Bohr}.   

As argued earlier, the influence of visco-elasticity is seen for low tape mass, a situation where the solutions are close to that of the differential algebraic equations \cite{Rumi05}. The orbits jump between the two branches of the peel force function almost instantaneously and therefore the peel velocity changes abruptly.  Thus, our equations are very stiff and therefore demand high accuracy in computation. For this reason, we calculate only the largest Lyapunov exponent. (Note that  if LLE is  positive, the system is chaotic.)  Indeed, under these conditions, even evaluating the LLE turns out to be time consuming due to slow convergence of the Lyapunov exponent. 

Given that equations of motion are chaotic, the equations are evolved till the phase space orbit settles down on the attractor.  The  method of calculating LLE involves choosing two neighboring trajectories $\vec{\xi}_i$ and $\vec{\xi}_j$  with an initial separation $d_{ij}(0)$ and evolving them for a short time $\Delta t$ and, monitoring the distance $d_{ij}(\Delta t)$. ( Any acceptable norm can be used. The simplest choice is to take $d_{ij} = \sqrt{\sum_{p=1}^M (\xi_i^{(p)}-\xi_j^{(p)})^2 }$, where the sum is over $M$ components of the vectors.) The initial distance is taken to be small compared to the size of the attractor.  Then, $\frac{1}{\Delta t} log \frac{d_{ij}(\Delta t)}{d_{ij}(0)}$ reflects the rate of divergence of the orbits. The procedure  is repeated by resetting the distance to $d_{ij}(0)$ along the direction of the evolved difference vector  $\vec{\xi}_i - \vec{\xi}_j$ so that the attractor is well sampled. Then, the largest Lyapunov exponent is given by 
\begin{equation}
\lambda = \frac{1}{K\Delta t} \sum_{1}^{K} log \frac{d_{ij}(\Delta t)}{d_{ij}(0)}. 
\end{equation}

\subsubsection{Power law distributions and scaling relations }

Given a fluctuating time series, the simplest statistical quantity that can be calculated is  the distribution of event sizes and their durations. Large number of driven systems exhibit power law distributions of event sizes and their durations. However, the definition of an event depends on the physical situation. Here, we use the magnitude of the local burst of the acoustic signal $\Delta R_{AE}$ as an event. This is defined as the  magnitude of the signal  from a maximum to the next minimum.  The corresponding time difference associated with $\Delta R_{AE}$ is taken as the duration $T$.  Then, the distributions $P(\Delta R_{AE})$ of the event sizes and  durations $P(T)$  follow a power law defined by   
\begin{eqnarray}
P(\Delta R_{AE}) &\sim & \Delta R_{AE}^{-\alpha},\\  
\label{event_two}
P(T) &\sim&  T^{-\beta}.
\label{duration_two}
\end{eqnarray}
In addition, event size and the lifetime are related though
\begin{equation}
\Delta R_{AE} \sim  T^x.
\label{lifetime}
\end{equation}

\begin{figure}[!b]
\vbox{
\includegraphics[height=3.6cm,width=7.0cm]{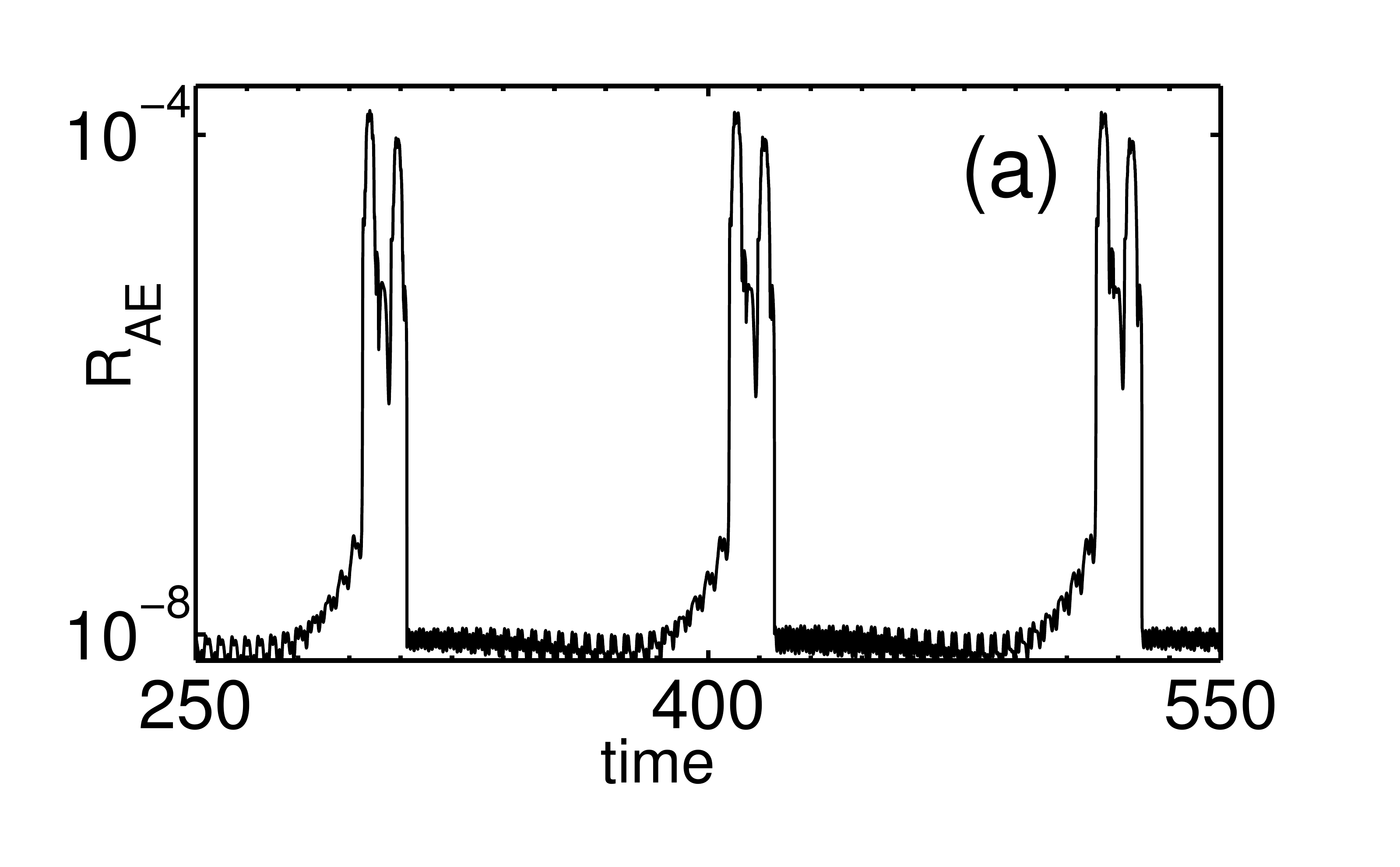}
\includegraphics[height=3.6cm,width=7.0cm]{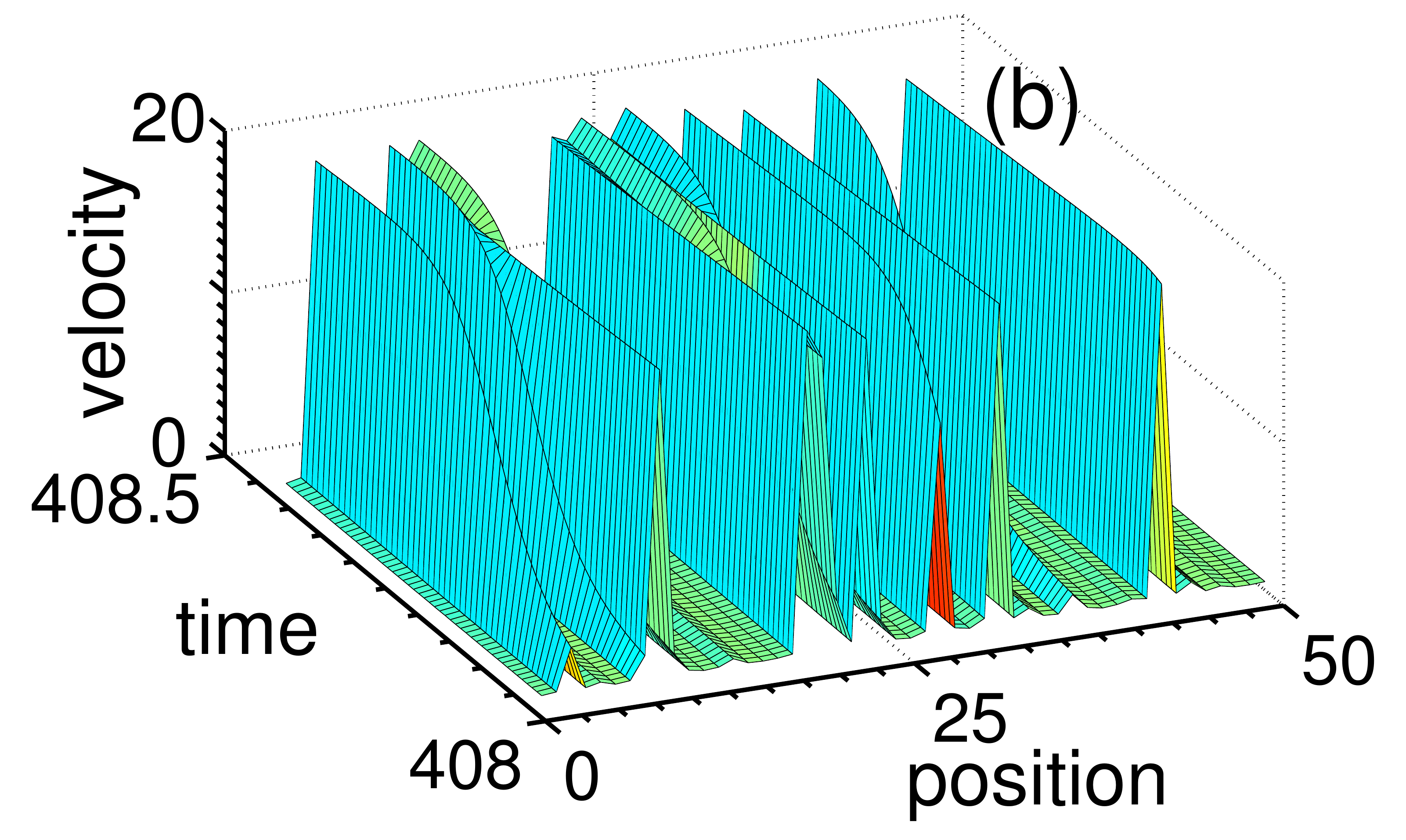}
}
\caption{(Color online b) Parameter values - $C_f = 7.88, m =10^{-3}, I=10^{-5}, V^s=1.48$, and $\gamma_u=0.01$. (a) Model acoustic energy plot for  the $k_0$-model. (b) Snapshot of stuck-peeled configurations for the $k_0$-model. 
}
\label{CHV1m3I5gu01_k0}
\end{figure}

\begin{figure}
\vbox{
\includegraphics[height=3.6cm,width=7.0cm]{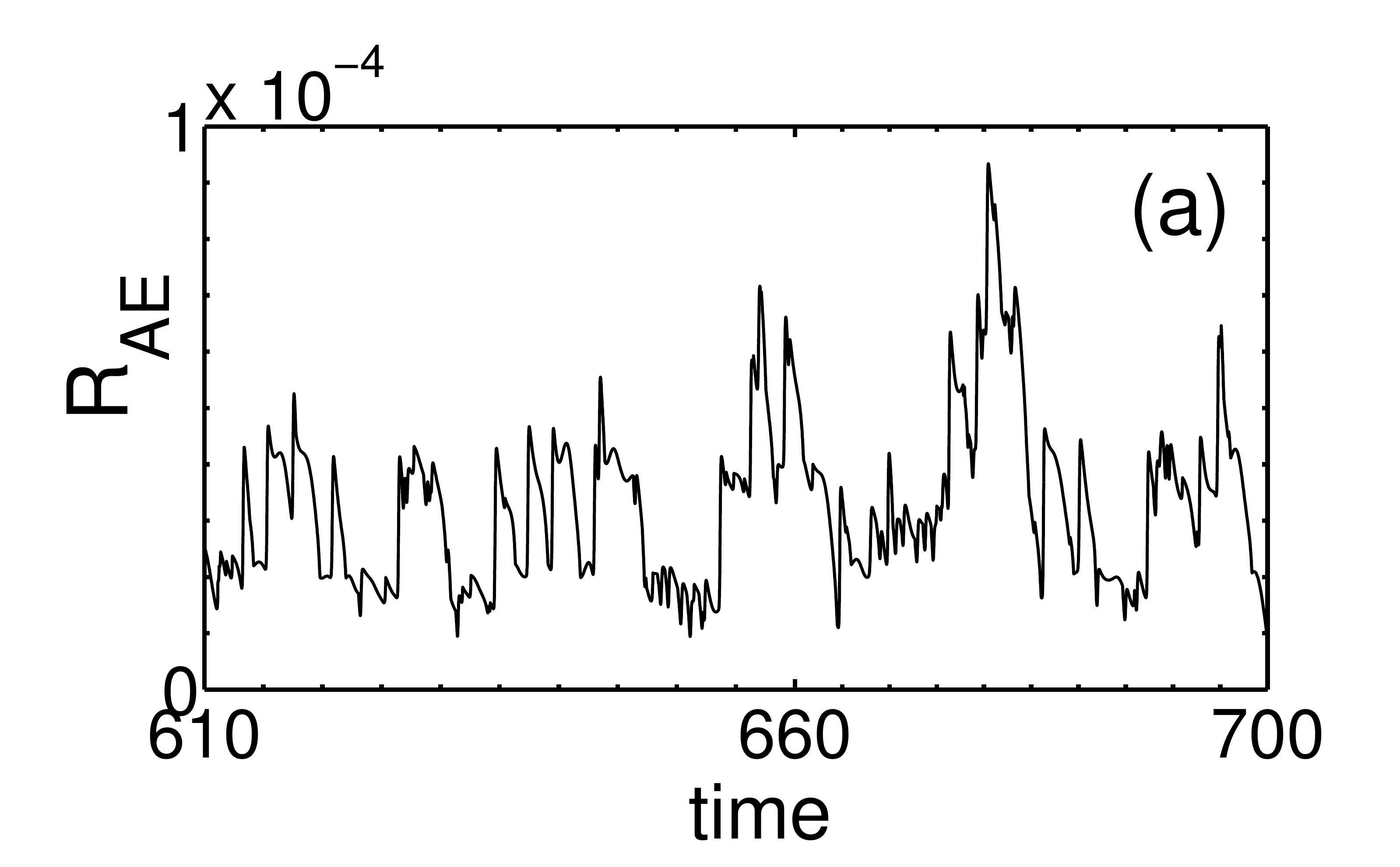}
\includegraphics[height=3.6cm,width=7.0cm]{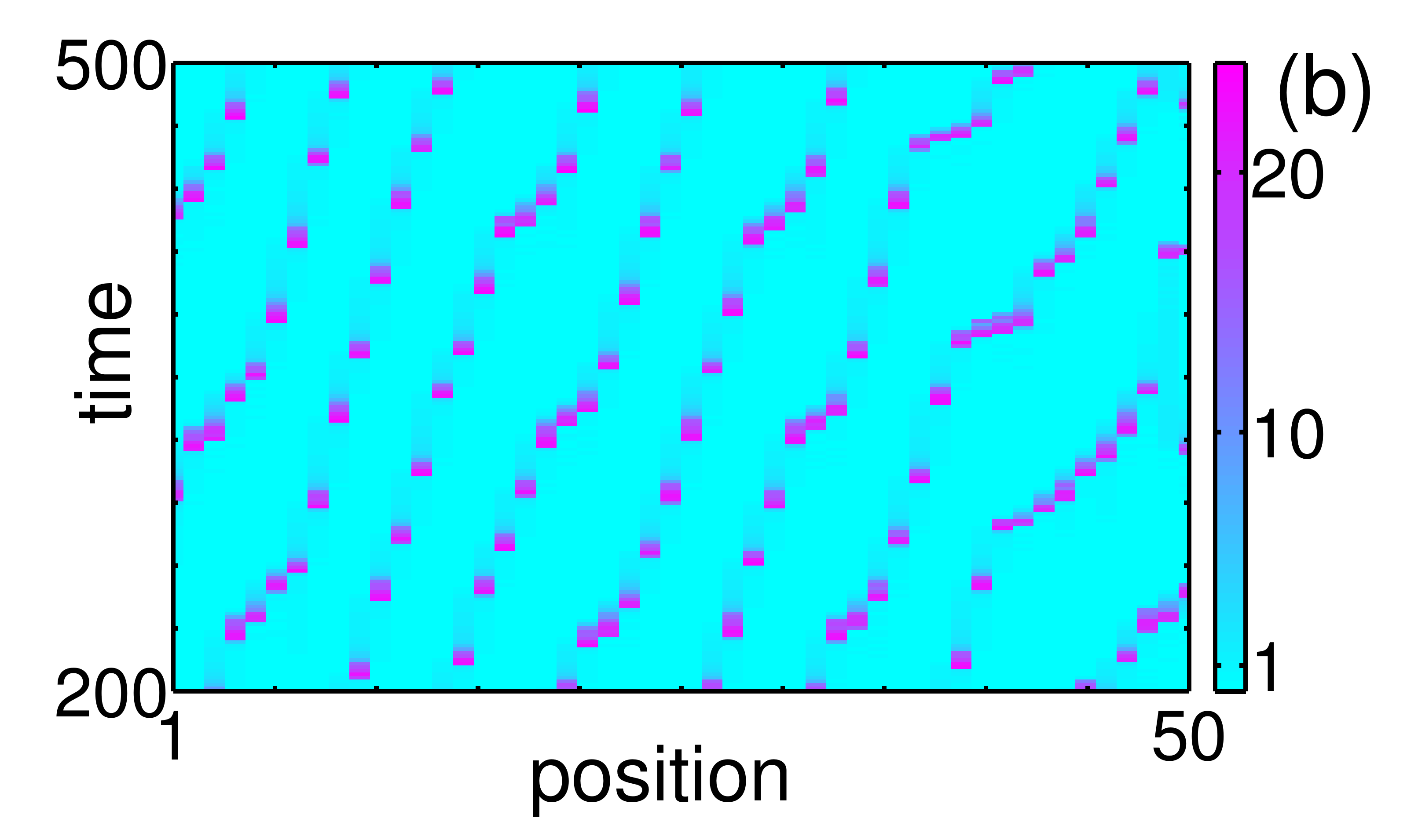}
\includegraphics[height=3.6cm,width=7.0cm]{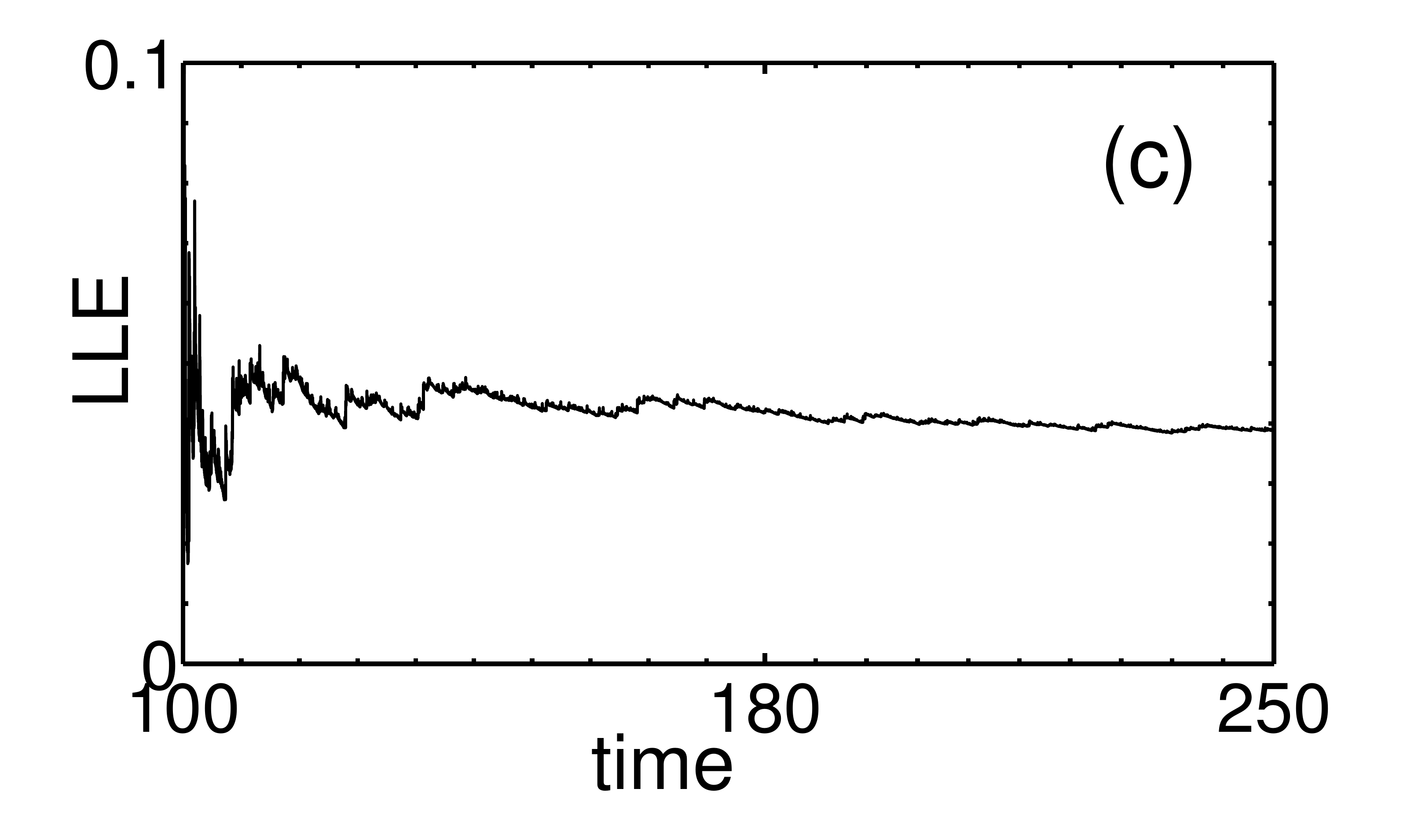}
\includegraphics[height=3.6cm,width=6.0cm]{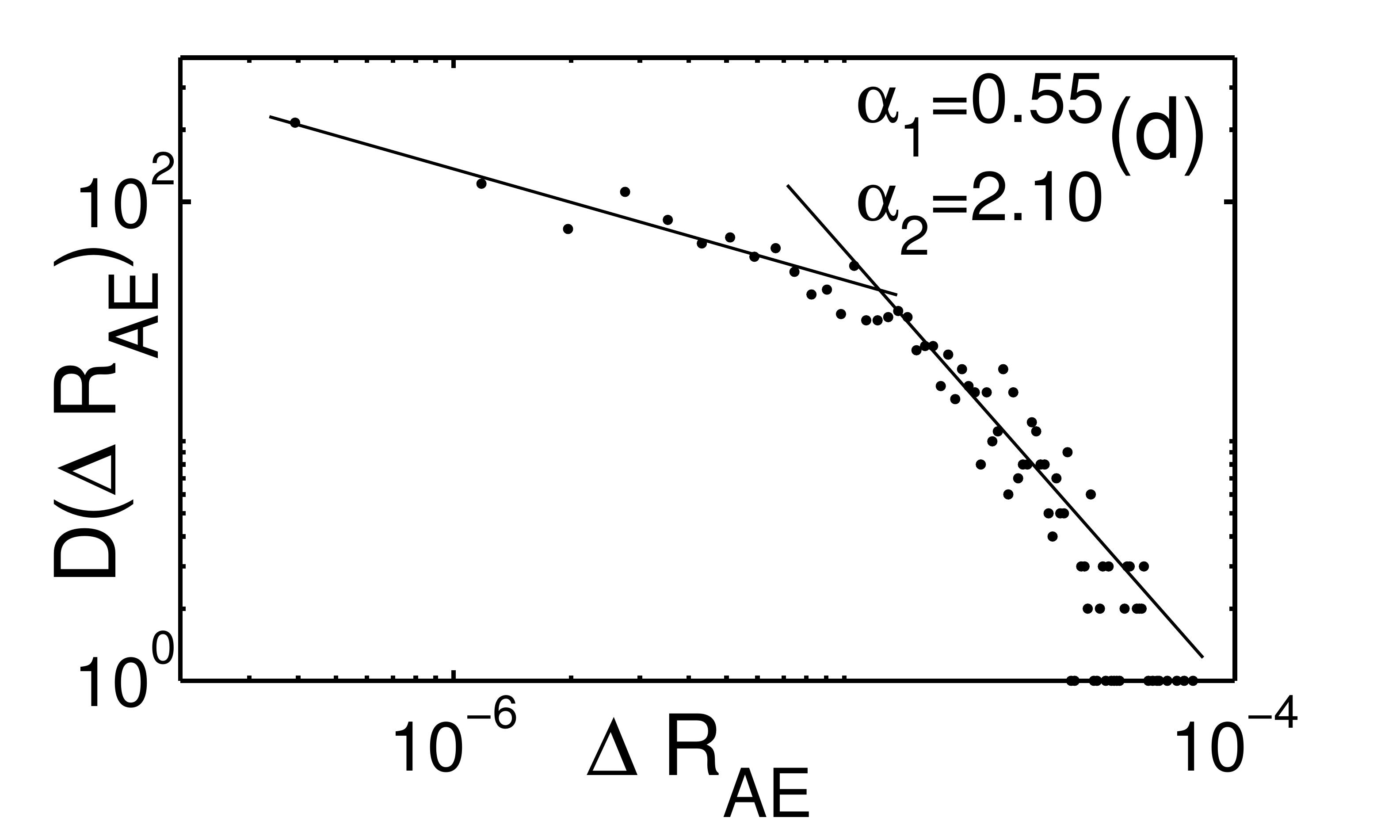}
\includegraphics[height=3.6cm,width=6.0cm]{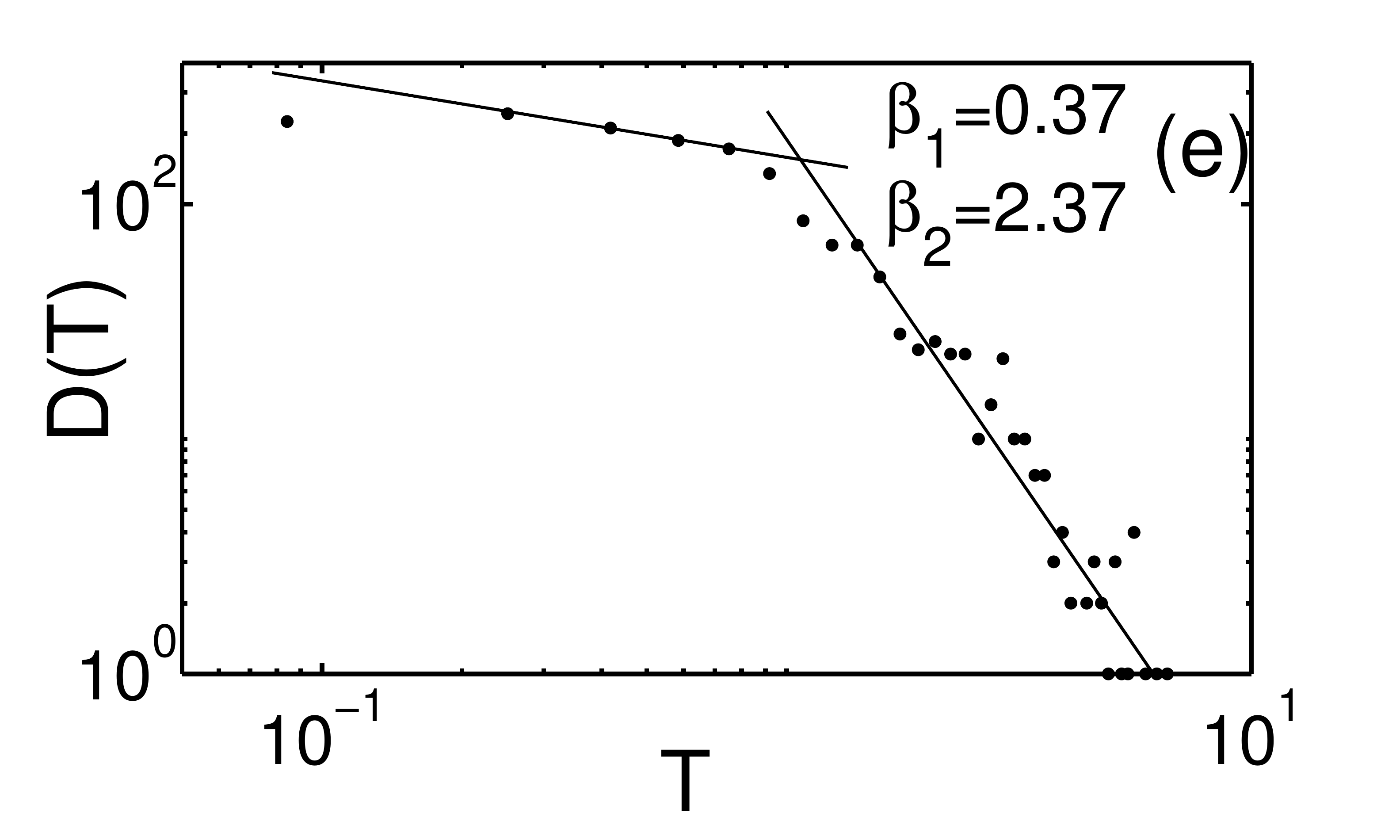}
\includegraphics[height=3.6cm,width=6.0cm]{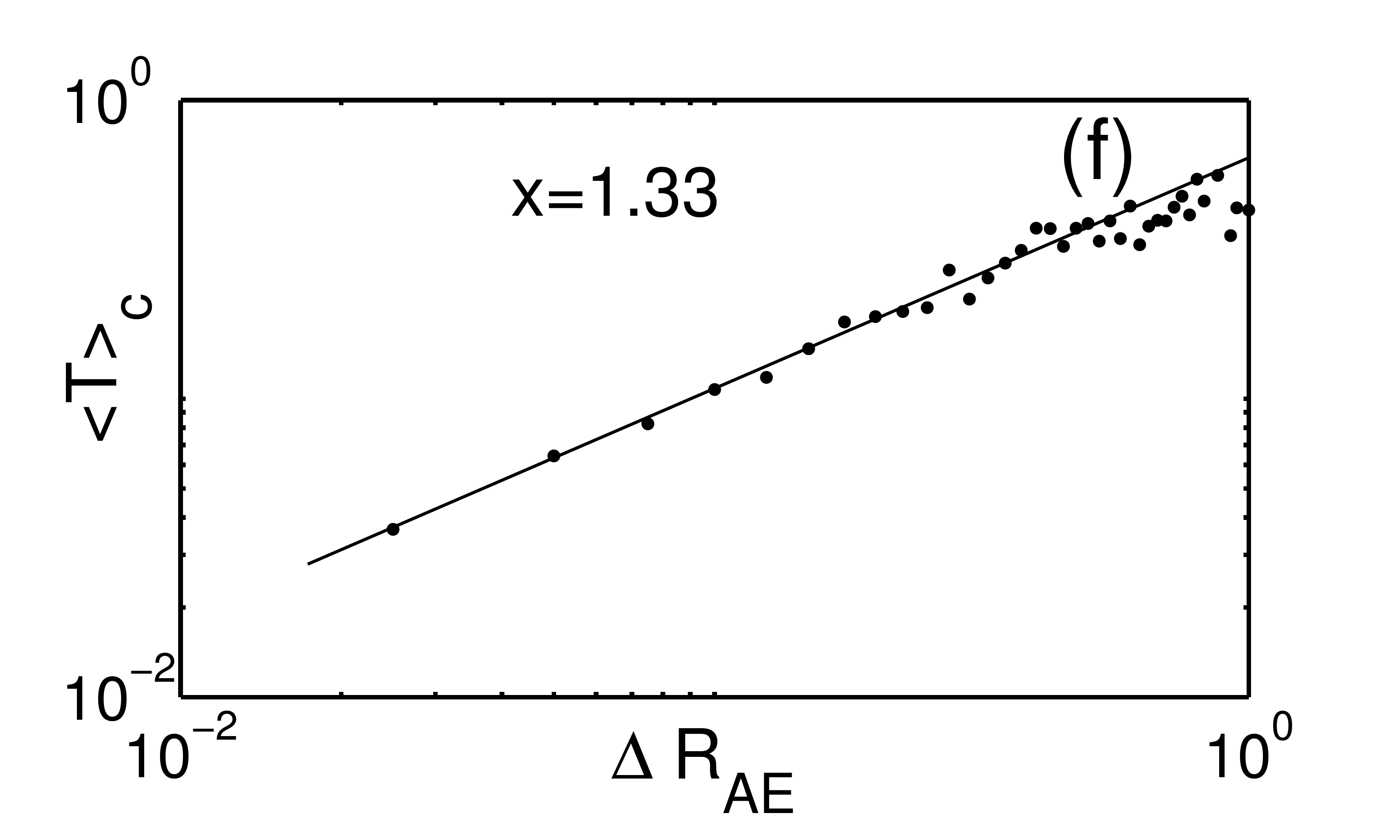}
}
\caption{(Color online b) Parameter values for the visco-elastic model - $C_f = 7.88, m =10^{-3}, I=10^{-5}, V^s=1.48, \gamma_u=0.01$  and $1/\tau_a = 1.9$ . (a)  Model acoustic energy plot for  the visco-elastic model.
(b) Time evolution of  the  stuck-peeled configurations. (c) Largest Lyapunov exponent for the visco-elastic model. (d,e) Two stage power law distribution for the event sizes $\Delta R_{AE}$ and their durations $T$. (f) Scaling relation between the event size $\Delta R_{AE}$ and the conditional average $<T>_c$. The exponent value is $x=1.33$.
}
\label{powerlaw}
\end{figure} 
However, our earlier investigations  have shown that  the distributions of event sizes follow either a single scaling regime or two distinct scaling regimes, one for small values of the variable and another for large values \cite{Rumi06,Jag08b}. For a two stage power law distribution, $\alpha=\alpha_1$ for small values of $\Delta R_{AE}$, and $\alpha=\alpha_2$ for large values of $\Delta R_{AE}$. Similarly, $\beta=\beta_1$ and $\beta_2$ for small and large values of $T$ respectively. We assume that there is a single scaling regime, with an exponent $x$, connecting the magnitude  of the event with its duration. While a scaling relation between the exponents has been derived for the case of a single stage power law distribution, a similar scaling relation is not available in the literature for the two stage power law distribution. Thus, our first task is to derive scaling relations valid for this case.  

Our derivation follows the approach due to R\'afols and Vives  \cite{Vives}. Using a joint probability distribution of event sizes and their durations $P(\Delta R_{AE}, T)$, we have derived scaling relations between the exponents in the appendix. The exponents corresponding to  small values of event sizes and durations are related through the standard scaling relation given by 
\begin{equation}
x(1-\alpha_1) = 1-\beta_1.
\label{smallevent}
\end{equation}
Surprisingly,  the exponent for the event size $\Delta R_{AE}$  corresponding to the second scaling regime ( i.e.,  large values) is completely determined by $\beta_1$ of the first given by
\begin{eqnarray}
\label{largevent}
x(\alpha_2 -1) &= & \beta_1 + 1,\\
\beta_2 &= & \beta_1 + 2.
\label{largeduration}
\end{eqnarray}

A few general comments are desirable. First, very often the statistics of large events are poor that  may  overshadow the possible existence of a power law for large values. This limitation applies even to  model systems let alone experiments. Second, in general, the statistics of event durations is known to be poor even in model systems. Thus, very often, it may not be possible to verify if the scaling relations are obeyed. Lastly, due to numerical accuracies, scaling relations are satisfied only approximately. 

\section{Influence of Visco-elastic Contributions to Peel Front Dynamics}
\label{sec3}

Here, we present a few representative results  where-in the influence of visco-elasticity is substantial and interesting. Henceforth, we refer to the present model (Eqs. (\ref{localconstraint}), (\ref{seqalpha}) and (\ref{sdotv1})) as the visco-elastic model and our earlier work in \cite{Rumi06,Jag08a,Jag08b} as the $k_0$-model. Unless otherwise stated, all other parameters are the same when a comparison is made. Here we investigate the influence of  the visco-elasticity of the adhesive on the peel front dynamics for a range of values of $C_f$ and $V^s$.  (Other parameters are fixed at $R^s=0.35$, $l^s=3.5 $, $k_t=1000N/m, 1/\tau_a=1.9$ and $N=50$ in units of the grid size). 

\subsection{ Case 1 : $C_f = 7.88$ }

For this case, the range of values of $(m,I)$ are  $(0.1,10^{-3}), (0.01,10^{-4})$ and  $(0.001,10^{-5})$. Since the effect of visco-elasticity is minimal for $m=0.1$, we will not discuss this case.

\subsubsection{Case 1(i): $C_f = 7.88$ and  $m=10^{-3},I= 10^{-5}$}

This case corresponds to low inertia of the tape and low inertia of the roller. As we shall see, this is also the case where there is a substantial change in the peel dynamics of the visco-elastic model compared to the $k_0$-model \cite{Jag08b}. Note also that $m=0.001$  corresponds to  vanishing tape inertial time scale and hence the phase space orbit jumps abruptly  between the two stable branches of the peel force function  $\phi(v^s)$.  For these parameter values,  burst type AE signals are seen at low pull velocity $V^s=1.48$ for the $k_0$-model as shown in Fig. \ref{CHV1m3I5gu01_k0}(a).  Recall that in our earlier work on the $k_0$-model, we had established a correspondence between the nature of the model acoustic energy and the sequence of peel front configurations responsible for the acoustic signal \cite{Rumi06,Jag08b}. The burst type of $R_{AE}$ arises when the system jumps between rugged configuration  that lasts substantial amount of  time and stuck-peeled configurations that last for a short time ( Fig. \ref{CHV1m3I5gu01_k0}(b)). 

In contrast, when the visco-elastic contribution is included, $R_{AE}$ turns  noisy and irregular as shown in Fig. \ref{powerlaw}(a), although there is a periodic component corresponding  to burst type signal of the $k_0$-model.   The noisy nature of $R_{AE}$ arises from the system traversing through a sequence of stuck-peeled configurations  that mostly contain only a few  peeled segments whose location keeps changing rapidly. (In this case, there are  much fewer stuck-peeled segments compared to  Fig. \ref{CHV1m3I5gu01_k0}(b).) The time evolution of the SP configurations is shown as a color plot in Fig. \ref{powerlaw}(b). As can be seen, the pattern appears to propagate to the right with a  well defined mean velocity. Considering the stuck-peeled segment as a double kink, it appears that the propagation is very similar to the kink propagation.  The nature of spatio-temporal patterns of the peel front and their  temporal evolution can be quantified by calculating the largest Lyapunov exponent from the system of equations. Figure \ref{powerlaw} (c) shows that the exponent value converges to 0.04. Thus these  equations are spatio-temporally chaotic for these parameter values. 

The statistics of $R_{AE}$ for the visco-elastic model shown in Fig. \ref{powerlaw}(a)  are analyzed in terms of the distributions of the event sizes  $\Delta R_{AE}$ and their durations $T$. ( Note that the statistics are given in terms of unnormalized distributions denoted by $D$ instead of normalized distribution $P$.) We have calculated the distribution of $\Delta R_{AE}$ and their durations $T$ denoted by $D(\Delta R_{AE})$ and $ D( T)$ respectively.  A plot of the event size distribution $D(\Delta R_{AE}) \sim \Delta R_{AE}^{-\alpha}$ exhibits a two stage power law as shown in Fig. \ref{powerlaw} (d).  For small values of  $\Delta R_{AE}$, the exponent value $\alpha=\alpha_1$ is close to $\alpha_1 \sim 0.55 \pm 0.01$, while for large values of $\Delta R_{AE}$, the exponent $\alpha=\alpha_2 \sim 2.1 + \pm 0.1$.  Similarly, the distribution of the duration of the events $ D(T)$ also exhibits a two stage power law with exponents $\beta_1=0.37 \pm 0.01$ and $\beta_2=2.37 \pm 0.05$ respectively for small and large values of $T$ (Fig. \ref{powerlaw} (e)). Further, as can be seen from Fig. \ref{powerlaw} (f), the event size $\Delta R_{AE}$ scales with  a {\it single exponent $x$ } with the conditional average $<T>_c$ (as assumed in our derivation )  with $x=1.33 \pm 0.07$. 

It is easy to check that the  scaling relations given by Eqs. (\ref{smallevent},\ref{largevent}) and \ref{largeduration} are satisfied quite closely. For instance,  for Fig. \ref{powerlaw}, the left hand side of Eq. (\ref{smallevent}) is $0.60$ while the right hand side is 0.63. Similarly, Eqs. (\ref{largevent})  is satisfied as $\alpha_2 =2.1$ which is numerically close to $ \frac{\beta_1+x+1}{x}\sim  2.03$. Further, $\beta_2 =2.37$ is found to be equal to  $\beta_1 + 2$.  We have indeed verified that the scaling relations are approximately satisfied in all cases where the statistics  are good.

\begin{figure}[b]
\vbox{
\includegraphics[height=3.6cm,width=7.5cm]{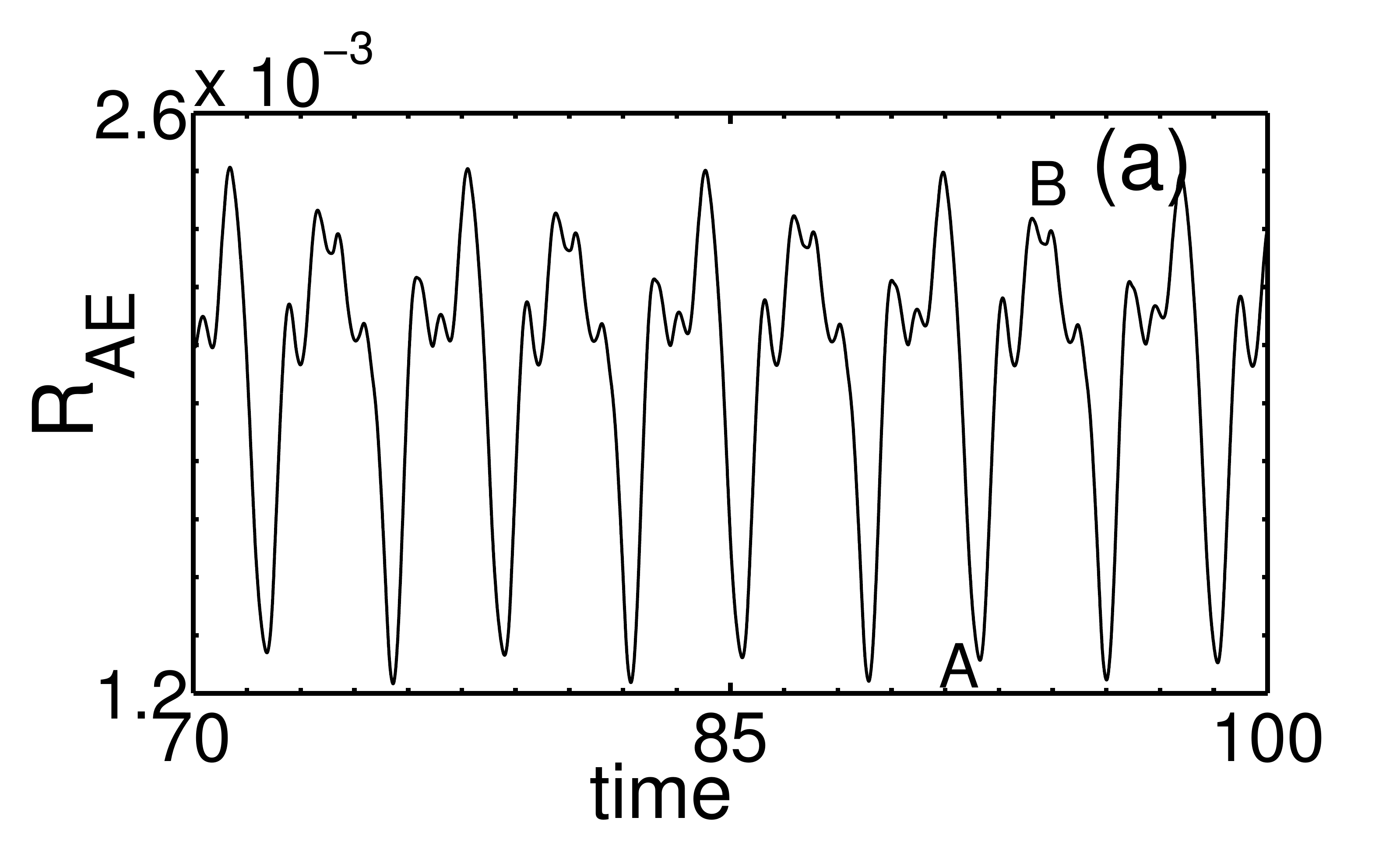}
\includegraphics[height=3.6cm,width=7.5cm]{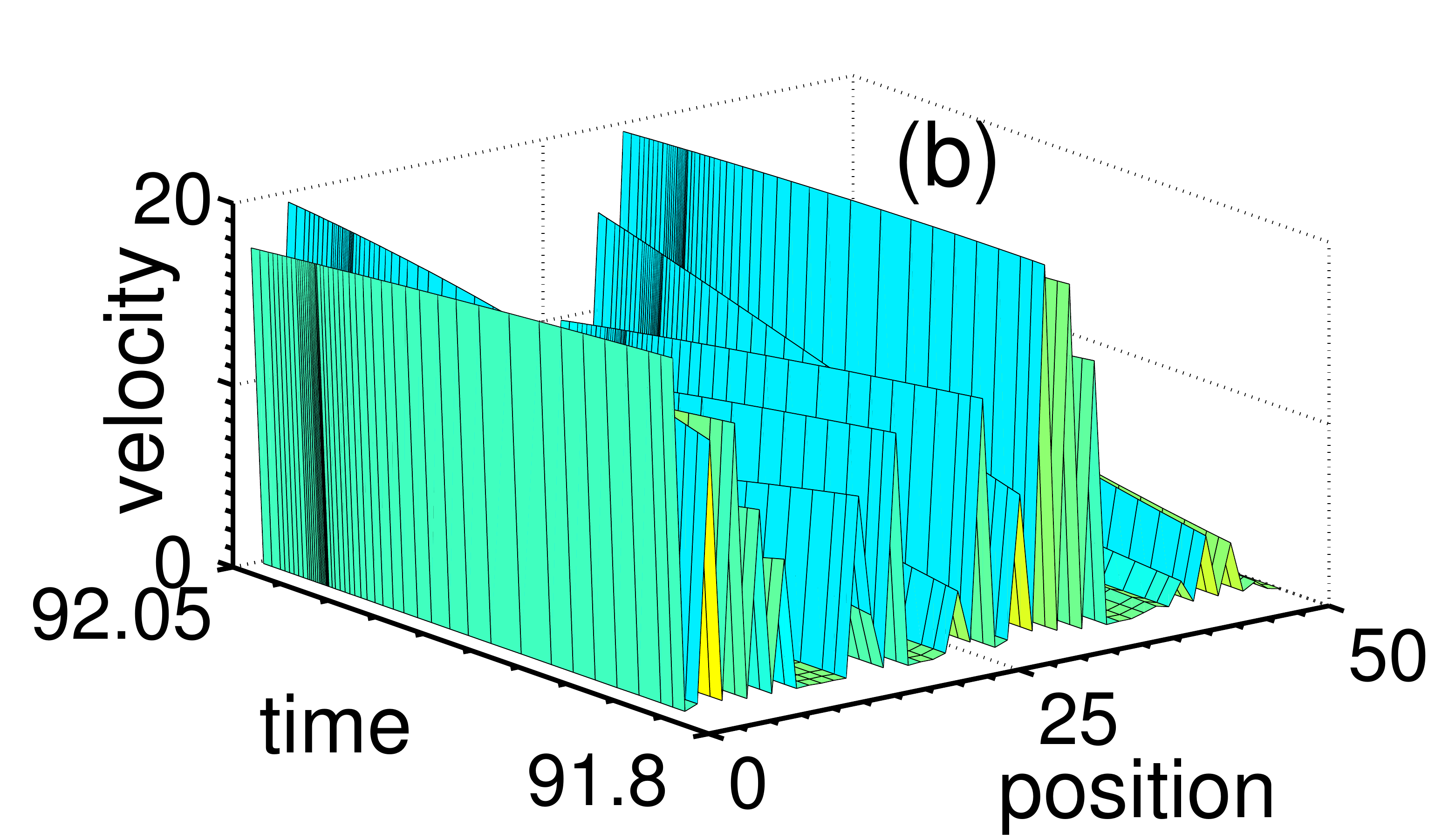}
\includegraphics[height=3.6cm,width=7.5cm]{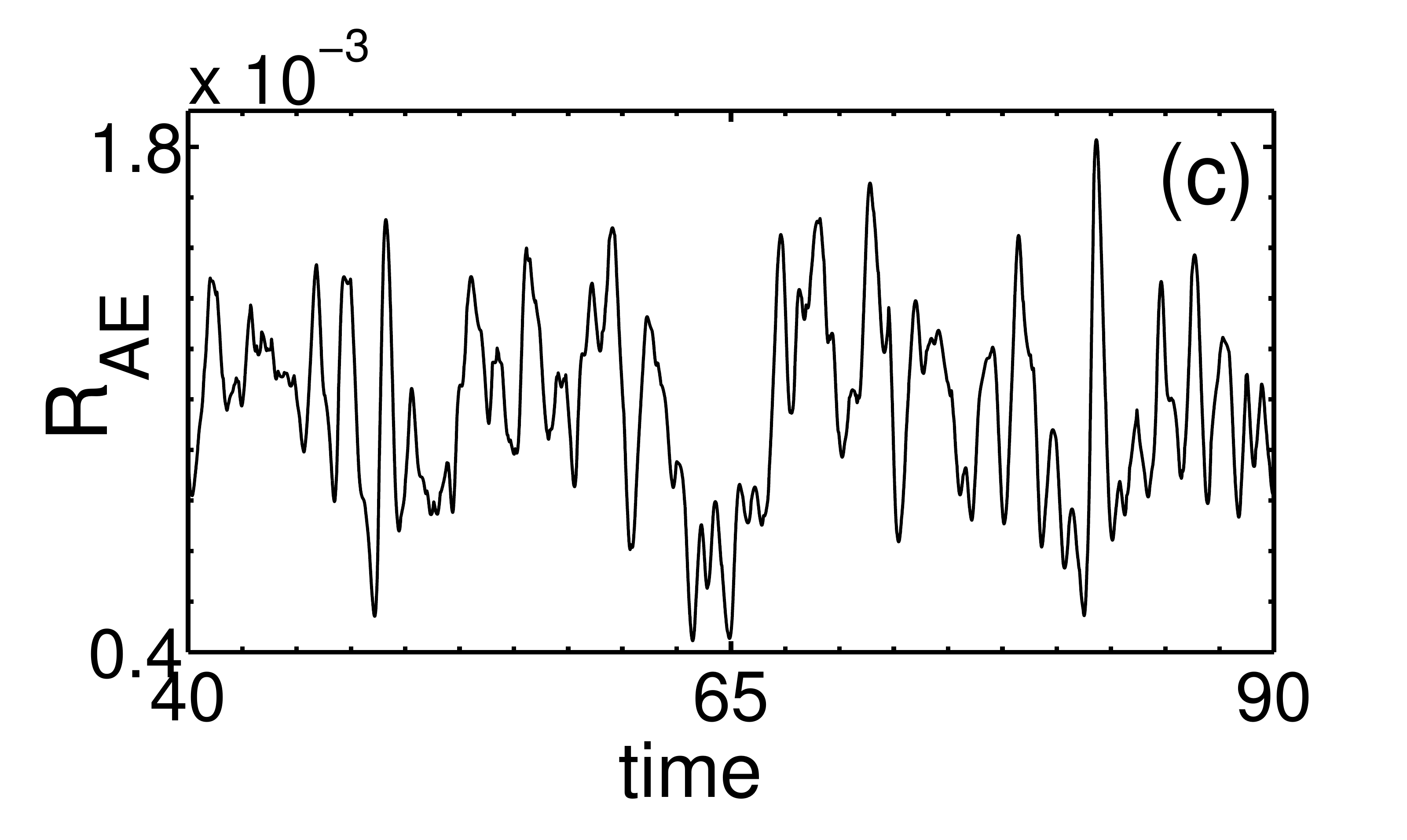}
\includegraphics[height=3.6cm,width=7.5cm]{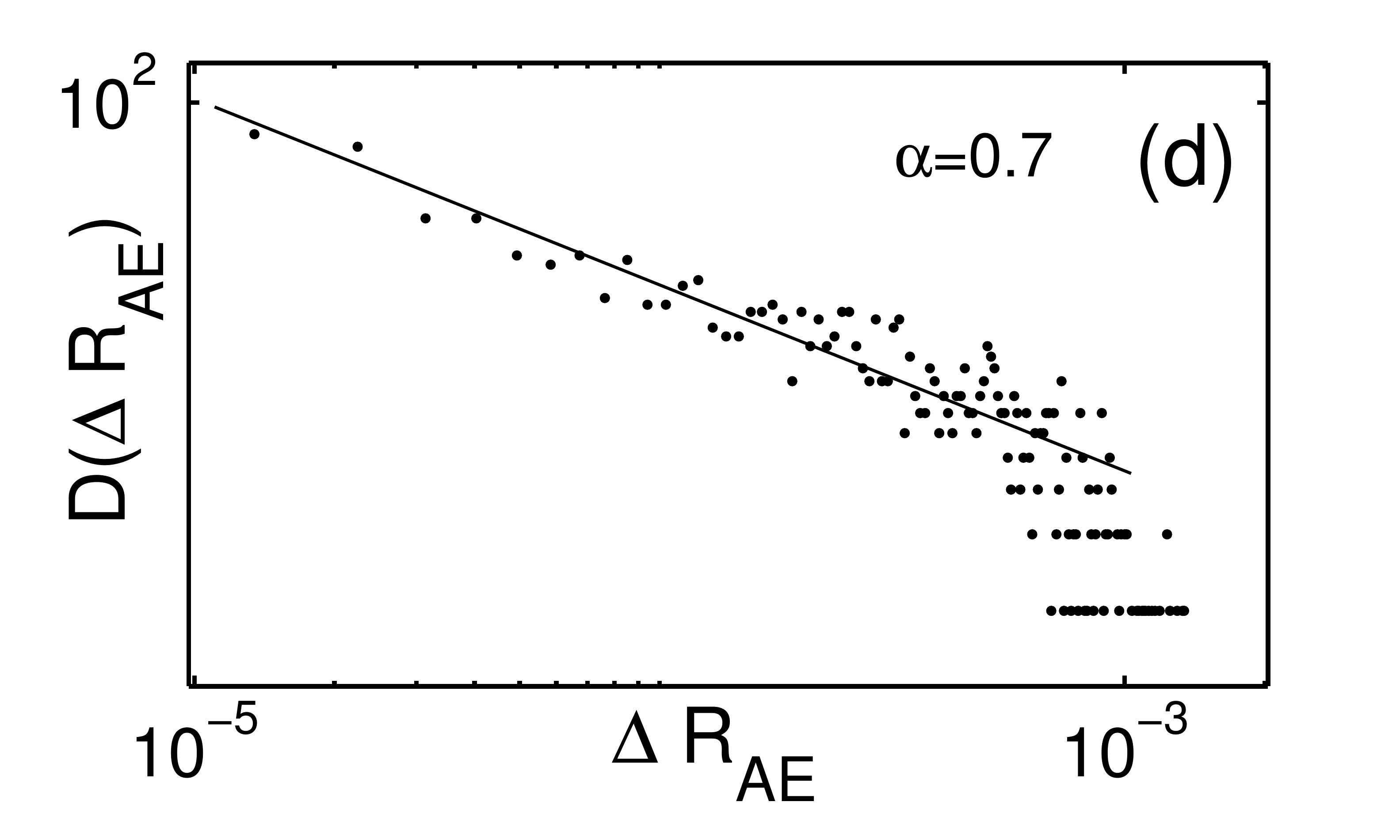}
}
\caption{(Color online b) Parameter values - $C_f = 7.88, m =10^{-2}, I=10^{-4}, V^s=4.48$, and $\gamma_u=0.01$. (a) Model acoustic energy plot for the $k_0$-model. (b) SP configuration for the $k_0$-model. Note the correspondence between the SP configuration shown and the point A on $R_{AE}$. (c) Model acoustic energy plot for the visco-elastic model. (d) Single stage power law for the visco-elastic model.}
\label{CHV4m2I4gu01_Visc}
\end{figure}

Now as we increase the velocity $V^s$, the general trend of the changes in the peel front dynamics for the visco-elastic model  are similar to those for  the $k_0$-model \cite{Jag08b} with minor  differences. The acoustic energy $R_{AE}$ is noisy  with a noticeable periodic component for  the $k_0$-model. The origin of the periodicity in $R_{AE}$ can be traced to the fact that the peel front goes through a repetitive sequence of SP configurations starting with a few stuck-peeled segments to a  maximum number. This overall periodicity  is less obvious in the acoustic signal $R_{AE}$ for the visco-elastic model. However, for $V^s=4.48$,  the distributions of amplitudes of the acoustic signal exhibit a two stage power law in both cases with nearly the same exponent values. But, the statistics of event durations is poor in both cases. For the $k_g$-model, the value of the LLE is $\sim 0.03$. Snapshot of SP configurations for the visco-elastic model are also  similar to those in Fig. \ref{CHV1m3I5gu01_k0} (b).

\subsubsection{Case 1(ii): $C_f = 7.88$ and  $m=10^{-2},I= 10^{-4}$}

For this case, for low pull velocity $V^s =1.48$, the peel front patterns for the present model are similar to those for the $k_0$-model \cite{Jag08b}. Only dynamic SP configurations are observed with the number of peeled segments changing continuously. (Plots of the SP configurations are not shown, as they are similar to other cases. See for example Fig. \ref{CHV1m3I5gu01_k0}(b).) Thus, the acoustic energy $R_{AE}$ is irregular with no trace of periodicity. 

However, as we increase $V^s$ to 4.48,  the AE signal for the $k_0$-model turns completely periodic. A typical plot  is shown in Fig. \ref{CHV4m2I4gu01_Visc}(a). Indeed, the phase plot in the $X-v^s$ plane is  periodic with a single loop. Even though only SP configurations are observed, they are long lived and are repetitive. Velocity-space-time plots corresponding to  maximum of  $R_{AE}$ marked $B$ is shown in Fig. \ref{CHV4m2I4gu01_Visc}(b). The minimum in $R_{AE}$ marked $A$ has fewer stuck-peeled segments compared to that for the point B (shown in  Fig. \ref{CHV4m2I4gu01_Visc}(b)).   In contrast, for the visco-elastic model, $R_{AE}$ remains irregular as shown in Figs. \ref{CHV4m2I4gu01_Visc}(c). The phase plot also appears to be chaotic. The largest Lyapunov exponent calculated from the equations of motion is $0.16$. The distribution function  $D(\Delta R_{AE}) \sim \Delta R_{AE}^{-\alpha}$ exhibits a single stage power law with an exponent $\alpha = 0.7 \pm 0.03$ as shown in Fig. \ref{CHV4m2I4gu01_Visc}(d). The exponent corresponding to event duration is $\beta = 0.5 \pm 0.02$ and that of $x= 1.82 \pm0.1$. It is clear $x( 1 - \alpha) = 0.55$ while $1- \beta =0.5$. Thus, the scaling relation Eq. (\ref{smallevent}) is well satisfied again. 

\subsection{ Case 2 : $C_f = 0.788$ }

For this  value of $C_f$, the four sets of values  of $(m,I)$ are:   $(10^{-1}, 10^{-2}), (10^{-2}, 10^{-3}),   (10^{-3}, 10^{-4})$  and $(10^{-4}, 10^{-5})$. However, here we report the results only for $(10^{-3}, 10^{-4})$ and  $(10^{-2}, 10^{-3})$ as the effect of visco-elastic contribution is not noticeable for high tape mass case while $m=10^{-4}$ is similar to $m=10^{-3}$ case.  
\begin{figure}[b]
\vbox{
\includegraphics[height=3.6cm,width=7.5cm]{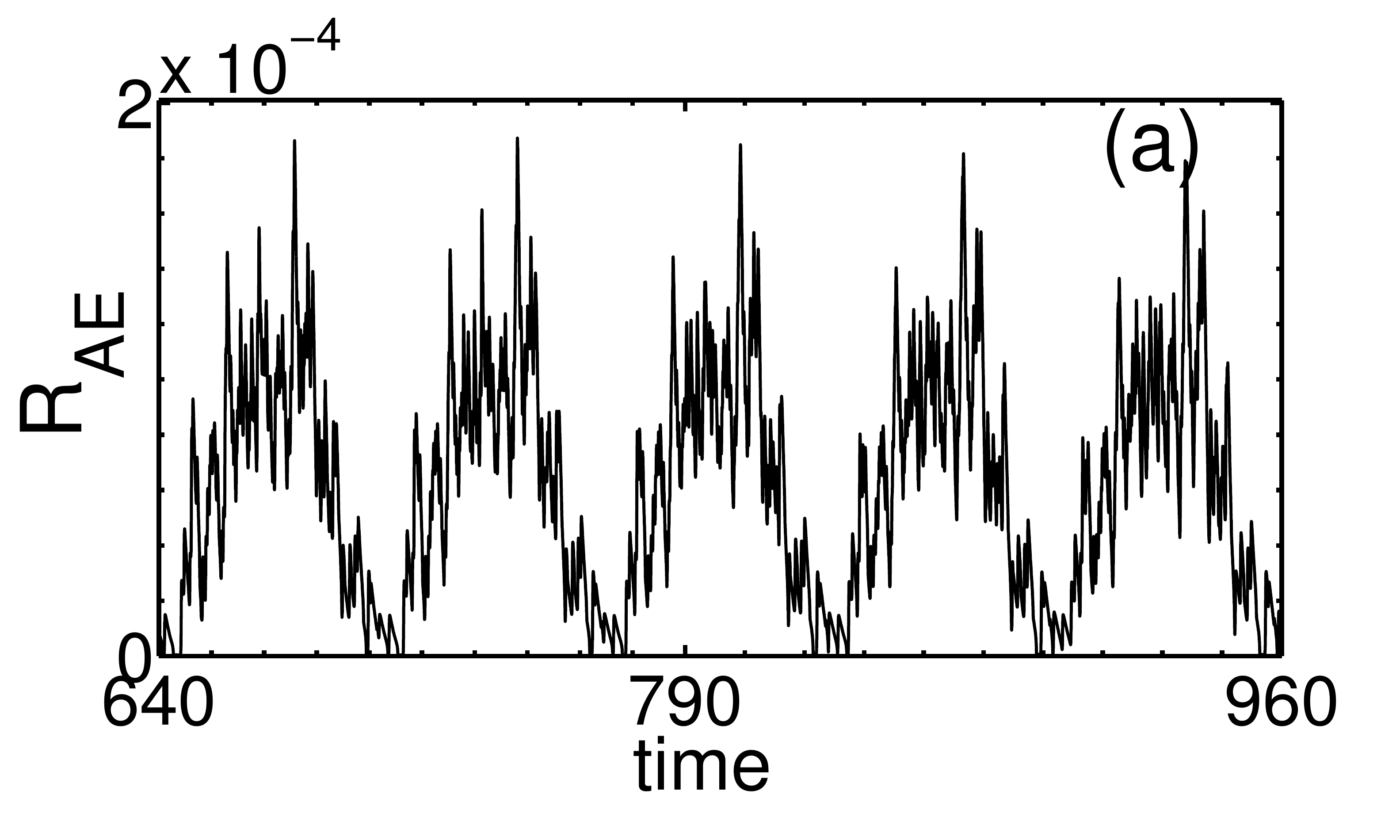}
\includegraphics[height=3.6cm,width=7.5cm]{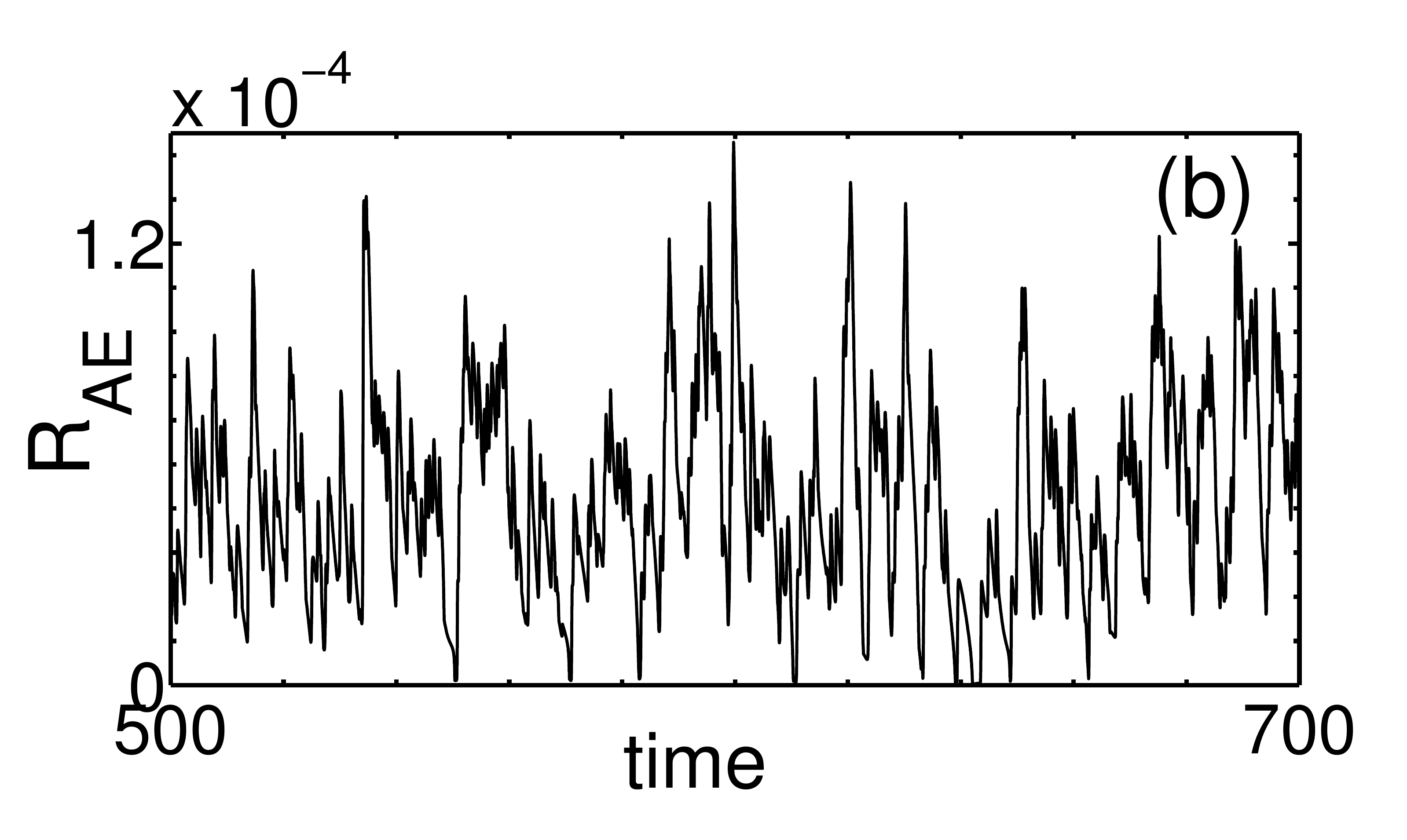}
}
\caption{Parameter values - $C_f = 0.788, m =10^{-3}, I=10^{-4}, V^s=1.48$, and $\gamma_u=0.01$ (a,b) Model acoustic energy plot for  the $k_0$-model and the visco-elastic model respectively. 
}
\label{CMV1m3I4gu01_Visc}
\end{figure}

\subsubsection{Case 2(i), $C_f = 0.788$, $m=10^{-3}$, $I= 10^{-4}$}

For this case also, there is a substantial change in the peel dynamics of the visco-elastic model compared to the $k_0$-model for $V^s=1.48$. For the $k_0$-model,  the model acoustic energy consists of a triangular envelope  of  rapidly fluctuating sequence of sharp bursts that repeats itself at near regular intervals as shown in Fig. \ref{CMV1m3I4gu01_Visc}(a). The peel process involves near periodic changes in the sequence of rapidly changing SP configurations starting with a single peel segment increasing to a maximum number of stuck-peeled segments, eventually reverting back to a single peel segment. (Figs. 9b and c of Ref. \cite{Jag08b} show the stuck-peeled configurations leading to the acoustic energy with a triangular envelope.)  The distribution $D(\Delta R_{AE}) \sim \Delta R_{AE}^{-\alpha}$ exhibits a two stage power law. For small values of  $\Delta R_{AE}$, the exponent value $\alpha = \alpha_1$ is close to $\alpha_1 \sim 0.5 \pm 0.02$, while for large values of $\Delta R_{AE}$, the exponent $\alpha=\alpha_2 \sim 2.0 \pm 0.1$. Since the statistics of event durations are poor, it is not possible to verify the scaling relations in this case. 

In contrast to the $k_0$-model, for $V^s =1.48$, the model acoustic  signal $R_{AE}$ appears irregular yet retaining some periodic component (of triangular bursts for the $k_0$-model) shown in Fig. \ref{CMV1m3I4gu01_Visc}(b),  The corresponding peel front configurations involve dynamic SP configurations. The largest Lyapunov exponent is $\sim 0.027$. The distributions of event sizes and durations exhibits two stage power laws with exponent values $\alpha_1=0.6 \pm 0.02, \alpha_2=2.1 \pm 0.01$, $\beta_1=0.6 \pm 0.02, \beta_2= 2.5 \pm 0.1$ and $x=1.51$.  The scaling relation are satisfied quite well. 

\begin{figure}
\vbox{
\includegraphics[height=3.6cm,width=7.5cm]{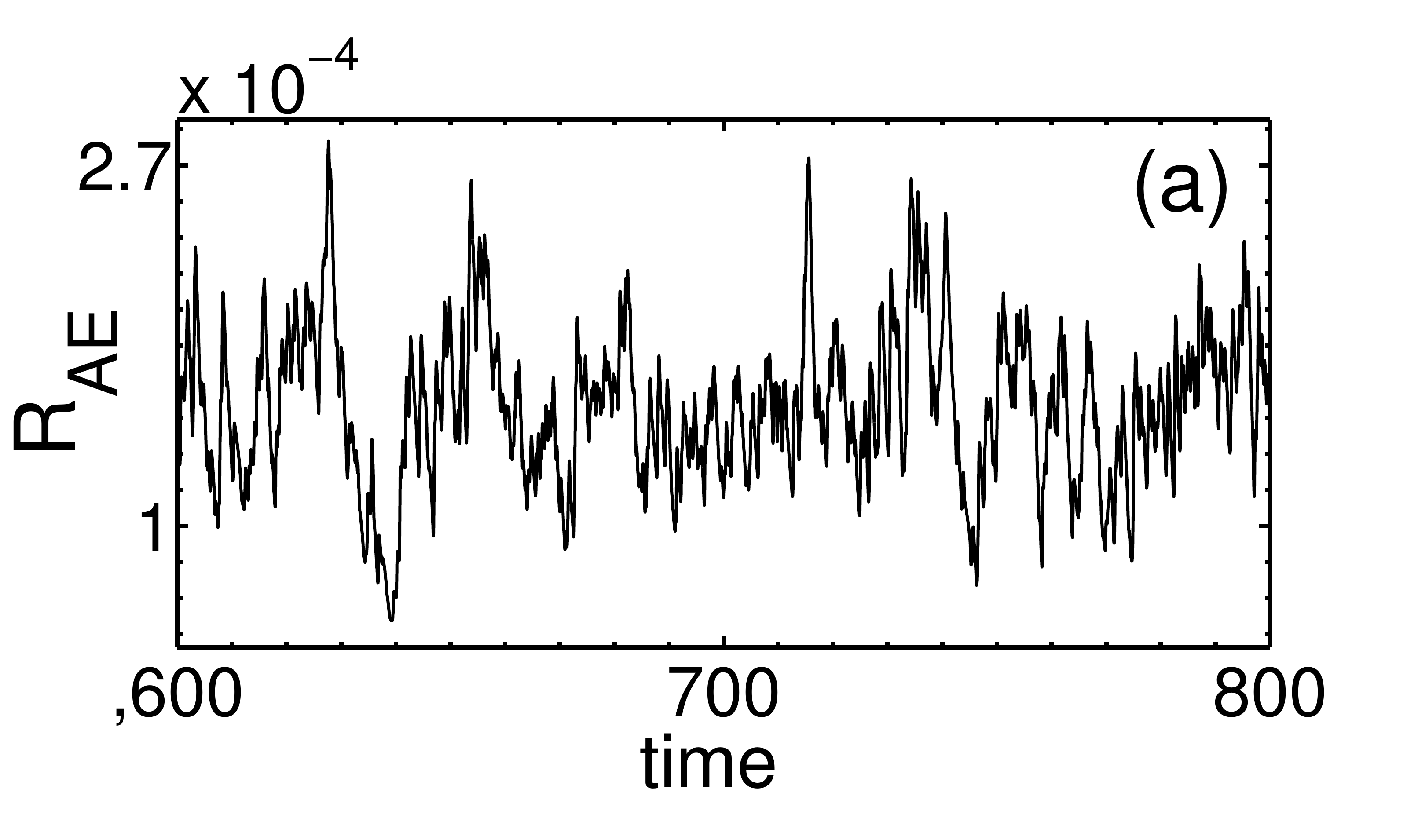}
\includegraphics[height=3.6cm,width=7.5cm]{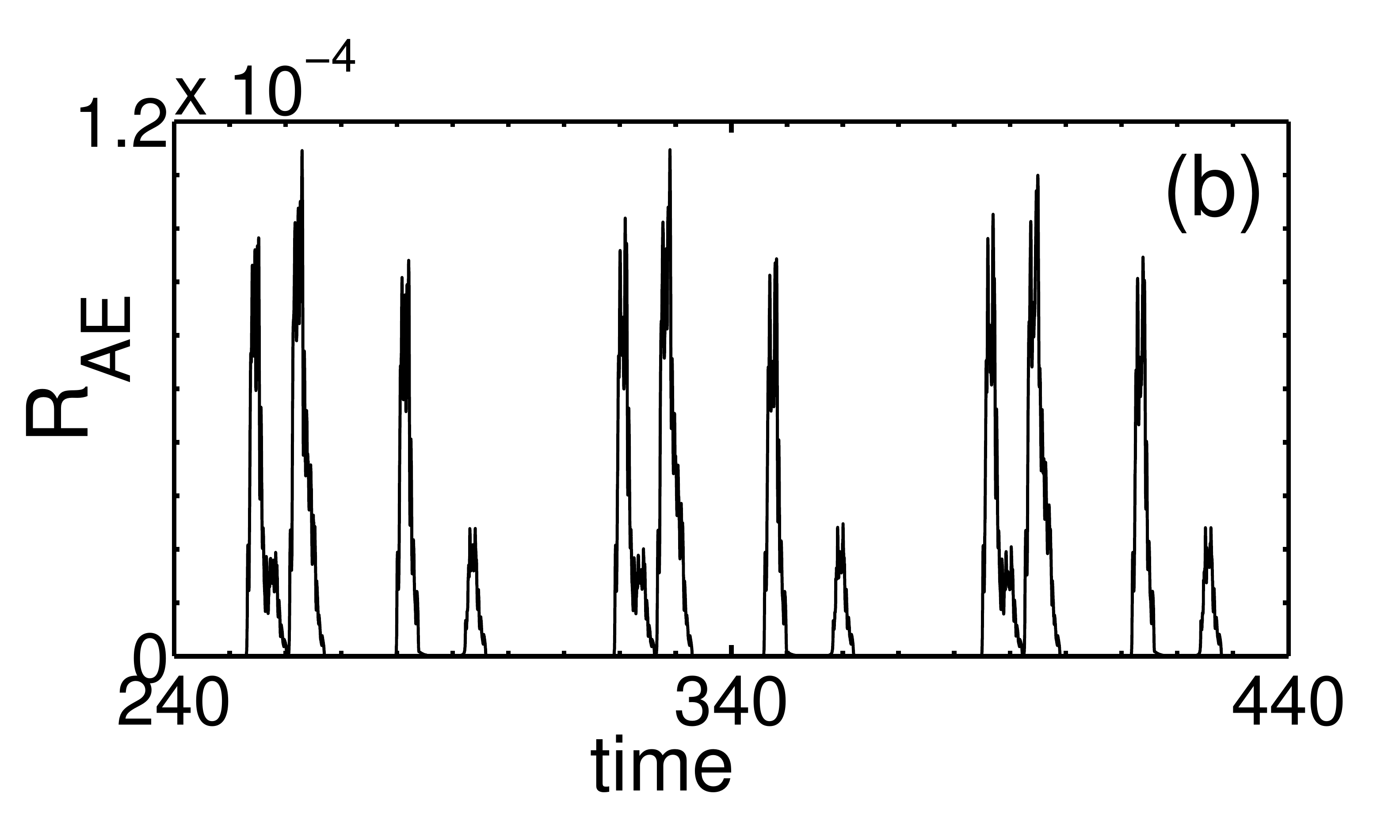}
}
\caption{(Color online b,c,d) Parameter values - $C_f = 0.788, m =10^{-3}, I=10^{-4}, V^s=4.48$, and $\gamma_u=0.01$ (a,b) Model acoustic energy plots for  the $k_0$-model and the visco-elastic model respectively. 
}
\label{CMV4m3I4gu01_Visc}
\end{figure}

As we increase $V^s$ to 2.48, the acoustic energy for the $k_0$-model becomes irregular with a noticeable superposed periodic component.  Further increase in the pull velocity to $V^s =4.48$ transforms $R_{AE}$ irregular without any trace of periodicity as shown in Fig. \ref{CMV4m3I4gu01_Visc}(a). Concomitantly, only dynamic SP configurations are seen.

For the visco-elastic model,  as we increase $V^s$, $R_{AE}$ still remains irregular for  $V^s =2.48$. A further increase to 4.48, the model acoustic emission signal turns out to be burst type with the bursts appearing at near regular intervals as shown in  Fig. \ref{CMV4m3I4gu01_Visc}(b). However, the nature of the bursts are  clearly different from Fig. \ref{CHV1m3I5gu01_k0}(a). In this case, the quiescent regions of  $R_{AE}$ correspond to configurations that are nearly smooth. Each burst in $R_{AE}$ is caused by the system jumping from this configuration to a sequence of rapidly varying SP configurations with only a few stuck-peeled segments. The largest Lyapunov exponent is close to zero suggesting the the equations are non-chaotic for the set of parameter values. The distribution $D(\Delta R_{AE}) \sim \Delta R_{AE}^{-\alpha}$, shows a two stage power law distribution  $\alpha_1=0.5 \pm 0.02$ and $\alpha_2 = 2.05 \pm 0.1$. However, the statistic of the event durations 
is poor and the distribution of the vent sizes shows no scaling regime.

Finally, some comments are warranted on the two contrasting dynamical responses of the peel front discussed above when the visco-elastic contribution is included.  In case 1(i) ($C_f=7.88$, $m=10^{-3}$ and $I= 10^{-5}$), at low pull velocity $V^s=1.48$,  burst type model acoustic energy $R_{AE}$ is observed for the $k_0$-model that changes over to irregular type with the addition of visco-elastic contribution.  For the $k_0$-model, on increasing the pull velocity, $R_{AE}$ exhibits an irregular form with a superposed periodic component, while $R_{AE}$ retains the totally irregular form for the visco-elastic model. In contrast, for the case 2(i) ($C_f = 0.788$, $m=10^{-3}$ and $I= 10^{-4}$), for the $k_0$-model,  $R_{AE}$ consists of a triangular  envelope of bursts for $V^s=1.48$ that transforms to aperiodic pattern at high velocity $V^s=4.48$. With the addition of the visco-elastic contribution, $R_{AE}$ is aperiodic at low pull velocity ($V^s=1.48$)  changing over  to burst type for $V^s=4.48$. Thus, the combined influence of visco-elasticity and pull velocity is contrasting in these two cases.

To understand this, consider the various mechanisms that contribute to the growth and decay of peel front instability. Recall that for a  continuous aperiodic type of AE signal,  there will always be stuck and peeled segments at any given time, i.e., when a segment that is in the peeled state gets  stuck, at least one other segment that is in the stuck-state peels out so that there is a dynamic balance.  In contrast, for burst like AE signal that has a near periodicity, the entire peel front spends a finite time on the low velocity branch and a short time in the transient  stuck-peeled configurations. Thus, for converting a burst type of signal with the addition of visco-elastic term, spatial heterogeneity needs to be sustained while the opposite should happen (though at high pull velocity) when an aperiodic signal changes to burst type.

Some insight can be obtained by  examining the influence of different terms in Eq. (\ref{sdotv1}). The equation has three new nonlinear terms compared to the $k_0$-model (first three terms on the left hand side of the equation) that can contribute to changes in the peel velocity $v^s(r)$. In addition, the coefficient $k$ of the diffusive term $ {\partial^2 X \over \partial r^2}$  now depends on $X/v^s$. Consider the first term on the LHS of Eq. (\ref{sdotv1}). Noting that $X/v^s \tau_a $ is always greater than unity when the peel velocity is in the region of slow velocity branch of $\phi(v^s)$, and noting $\dot X \approx V^s -v^s$, this term (with a nonlinear coefficient $ ({\partial X \over \partial r})^2$) contributes to the growth of any  perturbation. Note that other terms contribute to the decay. Indeed, even the diffusive term $k(X/v^s) {\partial^2 X \over \partial r^2}$ has a tendency to smoothen out SP configuration. As these terms depend on  the pull  velocity $V^s$, the influence of the visco-elastic term can be estimated by calculating individual contributions from these nonlinear terms for low and high velocities. We find that the first term  amplifies fluctuations in the peel front velocity for the case 1(i) at low velocities. Thus,  SP configurations are favored.

In the case of case 2(i), at high pull velocity ($V^s=4.48$),  we find the diffusive term $k(X/v^s) {\partial^2 X \over \partial r^2}$, more than compensates for the presence of the nonlinear amplifying (first) term and therefore has a tendency to smoothen out SP configurations. Dropping any of the nonlinear terms does not alter the burst type AE signal. This also suggest that a choice of small value for $k_0$ in the original model (without the visco-elastic contribution) should also give rise to burst like $R_{AE}$. Indeed, we have verified that burst like AE are seen if we choose a small value of $k_0$ ($k_0=0.05$) in the $k_0$-model. While this discussion offers some understanding, the set of coupled equations are much too complicated for any further analysis.

\begin{table}[!h]
\caption{Statistical and dynamical invariants for the set of parameters where the influence of visco-elastic contribution is significant. The values in the first row correspond to the visco-elastic model  and that in the second row to the $k_0$-model. NC is nonchaotic.}

\label{Exptstat}
\centering
\begin{tabular}{p{2.6em} p{2.5em} p{2.5em} p{2.5em} p{2.5em} p{2.5em} p{2.5em}  p{3.5em} }  \hline
\hline
Model & $V^s$ &  $C_f$ & $m$ & $I$ &$\alpha_{1}$& $\alpha_{2}$ & LLE\\ \hline

$k_g$ & $4.48$ & $7.88$ & $10^{-2}$ &$10^{-4}$&0.70&  &0.160 \\ 
$k_0$ &   $4.48$     &  $7.88$   &  $10^{-2}$      &   $10^{-4}$   &  &  &NC\\ 
$k_g$ &{$1.48$}  & $7.88$ & $10^{-3}$ &$10^{-5}$&0.55&2.10&0.040\\
$k_0$ &{$1.48$}   &$7.88$   &$10^{-3}$ & $10^{-5}$  &  & &NC\\ 
$k_g$ &{$1.48$}  & $0.788$& $10^{-3}$ &$10^{-4}$&0.60&2.20&0.082\\
$k_0$ &{$1.48$}  &  $0.788$& $10^{-3}$   & $10^{-4}$ &0.50&2.00&0.270\\ 
$k_g$ &{$4.48$}  & $0.788$& $10^{-3}$ &$10^{-4}$&0.78&2.00&0.008\\
$k_0$ &{$4.48$}  & $0.788$& $10^{-3}$ &$10^{-4}$&0.75&  &0.350\\ \hline
\hline
\end{tabular}
\end{table}

A few systematics have been noted in our studies  on the  power law distributions of the event sizes and durations (for entire range  of parameter values). First, the distributions are  either a single stage power law or a two stage power law. Second, if it is single stage power law distribution, the exponent corresponding to the magnitude of the events $\Delta R_{AE}$  is invariably $\sim 0.7$. In contrast,  the two stage power law distributions are of two types. The exponent corresponding to  small values of $\Delta R_{AE}$  is typically $\sim 0.5$  while that corresponding to large values of  $\Delta R_{AE}$ is always close to $\sim 2.0$. The exponents corresponding to the duration of the events  are related though $\beta_2 \approx \beta_1 + 2$. The derivation of the scaling laws ( see appendix) provides some insight into these origin of the systematics. Table I summarizes the changes induced  with the addition of visco-elastic contribution. 

It must be stated that while the results given above only deal with the set of parameters where the dynamics changes substantially with the addition of visco-elasticity, there is a range of parameter values for which there are changes that are not as dramatic. 

\subsection{Summary of the model acoustic energy profiles and the associated  peel patterns}

\begin{figure}[b]
\vbox{
\includegraphics[height=3.6cm,width=8.5cm]{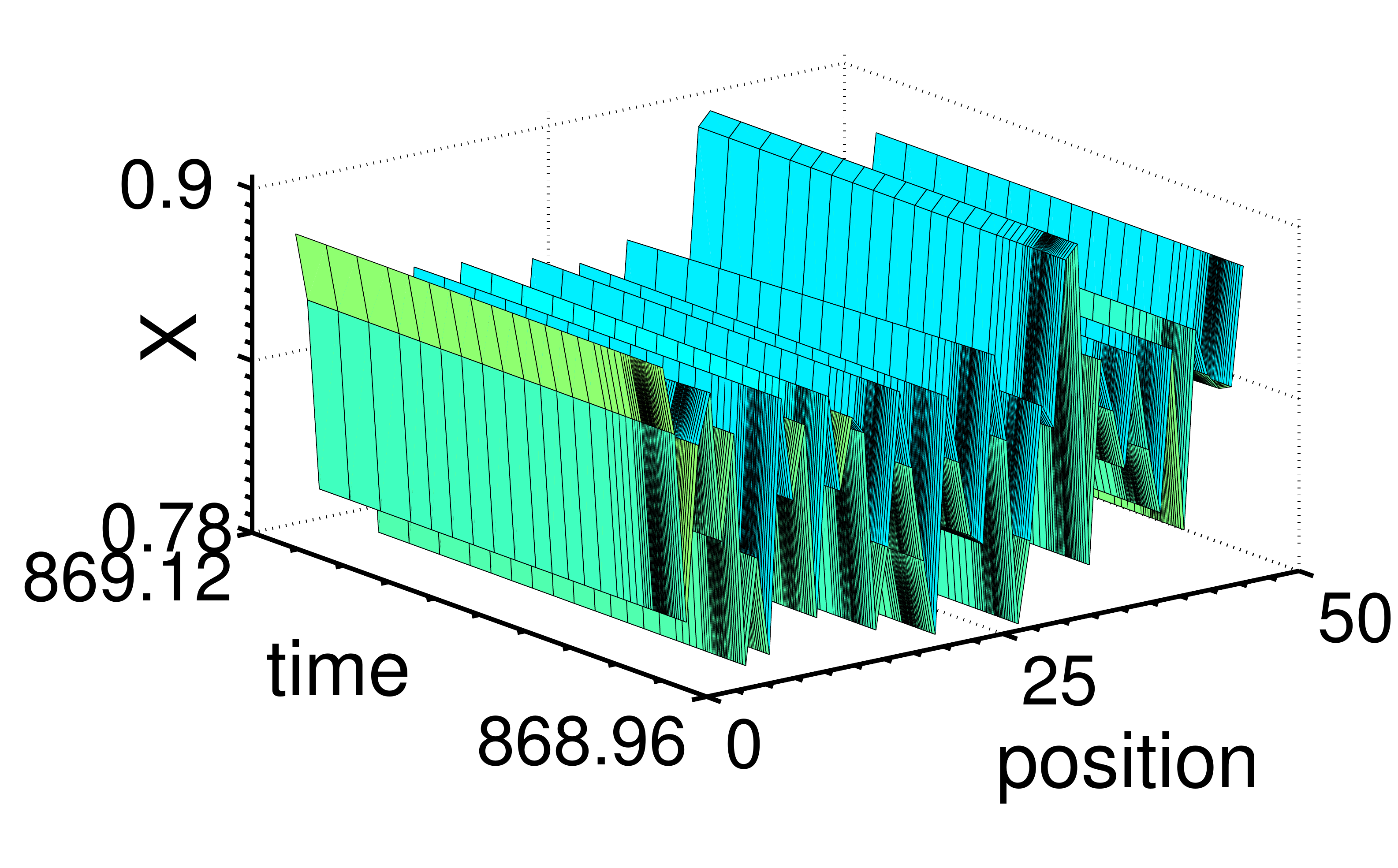}
}
\caption{(Color online) Spatial profile of the peel front  for the  $k_0$-model for $C_f = 0.788, m =10^{-3}, I=10^{-4}, V^s=1.48$, and $\gamma_u=0.01$.
}
\label{CMV1m3I4gu01_Visc_X1}
\end{figure}

All the peel front patterns observed in the $k_0$-model are also observed in the visco-elastic model.   These patterns can be classified as rugged, corrugated  and stuck-peeled configurations. Among the SP configurations, there are substantial variations, for example, rapidly  changing, long lived etc. The stuck-peeled configurations mimic the fibrillar pattern observed in experiments. A typical  model peel front profile shown in Fig. \ref{CMV1m3I4gu01_Visc_X1} can be compared with  the fibrillar pattern in \cite{Urahama,Dickinson,Yamazaki}. Despite the numerous possible configurations, {\it  only five different forms of the model acoustic energy $R_{AE}$ could be  identified} for the entire set of parameters space studied for both the visco-elastic model and $k_0$-model. This is surprising since $ R_{AE}(\tau)$ is the spatial average over the local displacement rate of all the allowed peel front configurations. Despite this a specific sequence of peel fronts configurations \cite{Rumi06,Jag08a,Jag08b} are found to be associated with each type of  the model acoustic energy $R_{AE}$. Here we list the five distinct  model acoustic  emission signals and the associated peel front configurations that generate the AE signals.

i) Type I:  Burst type AE pattern  arises when the entire peel front jumps from  a smooth or rugged configuration corresponding to the low velocity branch of $\phi(v^s)$ to a transient set of stuck-peeled configurations (see Fig. \ref{CHV1m3I5gu01_k0}(a) and \ref{CMV4m3I4gu01_Visc}(b)).

ii) Type II:   Irregular and continuous type of AE pattern is the most complex type. The chaotic nature of the AE  pattern can be identified with a set of rapidly changing set of SP configurations. The local minimum of $R_{AE}$ corresponds to fewer number of stuck-peeled segments compared to that at the following  maximum of $R_{AE}$. 

iii) Type III: Continuous irregular type AE signal with a noticeable periodic component is also associated with  the dynamic SP configurations. Here,  the peel front traverses through a nearly periodic sequence of  SP configurations starting with a few stuck-peeled segments to a maximum number. The minimum (maximum) in the nearly periodic profile of  $R_{AE}$ corresponds to SP configurations with a few (maximum) stuck-peeled segments.

iv) Type IV: Nearly periodic rapidly fluctuating convex envelope of AE bursts separated by a quiescent region are caused when the peel front traverses through a set of  SP configurations with increasing number of stuck-peeled segments starting with a rugged configuration. This type of signal is essentially type III, except that the number of bursts within one cycle is substantially more than that in type III. The quiescent region of  $R_{AE}$ corresponds to rugged configuration while the SP configuration with a maximum number of stuck-peeled segments to a maximum of $R_{AE}$.

iv) Type V :  Completely periodic AE signals are produced  when the peel front traverses through a periodic set of  SP configurations. The usual correspondence of the minimum (maximum) in $R_{AE}$ with the minimum (maximum) number of stuck-peeled segments holds.  

\section{Summary and Conclusions}
\label{sec4}

In summary, a detailed analysis of the peel front dynamics and the associated acoustic energy signal shows that the addition of visco-elasticity of the adhesive alters the dynamics. In particular, we have demonstrated that the influence of visco-elasticity is observed for low and medium tape mass.   The combined effect of the roller inertia and pull velocity makes the acoustic energy noisier for small tape mass and low pull velocities compared to the burst type emission for the $k_0$-model. For intermediate tape mass and roller inertia for high velocity, a periodic model acoustic energy signal of the $k_0$-model is transformed into an irregular pattern. In contrast, for low tape mass, intermediate roller inertia and high pull velocity, the original irregular acoustic energy signal is transformed to burst like with the addition of visco-elasticity. Despite the multitude of allowed spatio-temporal configurations, we find only five types of model acoustic emission signals among multitude of possibilities of the peel front configurations.  Of these, the most interesting is the stuck-peeled configurations. Even  among the SP configurations, there are substantial variations, for example, rapidly  changing, long lived, propagating  etc. Of these, Finally, the stuck-peeled configuration are interesting since they resemble the observed fibrillar patterns of the peel front. This is shown in Fig. \ref{CMV1m3I4gu01_Visc_X1}.

Two quantitative methods of analysis are introduced. The dynamical changes are quantified by calculating the largest Lyapunov exponent. Statistical features of the model acoustic energy signals are  analyzed by calculating the statistical distributions of the event sizes and their durations. Both single and two stage power law distributions are observed. Scaling relations between the exponents are derived, which show that the exponents corresponding to region of large values of event sizes and durations are completely determined by those for small values. The scaling relations are found to be satisfied in all cases studied where the statistics are satisfactory. 

By necessity, the work also addresses the conceptual problem of including visco-elastic effect of the adhesives applicable to intermittent peel front dynamics. This  has been done   within the context of $k_0$-model and thus the results of $k_0$-model form the basis for comparison. In our work on the $k_0$-model, the visco-elastic nature of the adhesive was included {\it only in an indirect way} by choosing the spring constant of the peel front to be small. This clearly ignores two important features, namely the time and rate dependence of the adhesive material. Further,  these properties are always measured in stationary deformation conditions that are {\it not applicable to intermittent flow}.  Thus, a major obstacle in accomplishing this objective is that there is no known method for including time dependence of visco-elastic material (the elastic modulus) and rate dependent deformation valid for unstable intermittent peel situations.  While the former is well described by Eq. (\ref{kg}), rate dependence arises from a subtle interplay of several internal relaxation mechanisms, and is certainly a complex phenomenon. In the context of the peel problem, the relevant physics that needs to be captured is that at low peel rates, the adhesive should behave like a viscous liquid while at high peel velocities, it should behave like a solid. Using the fact that all variables in the $k_0$-model already have a built-in  rate dependence on the pull velocity, we eliminate the explicit time dependence in favour of dynamical variables, here displacement and velocity of the peel front. Note that the algorithm combines these two distinct properties into a single equation.   

Interestingly, the  approach introduced is quite general and offers a platform for investigating rate dependent effects in other unstable dynamical situations.  For instance, one can adopt this method in generalizing the PLC model \cite{Kubin} where an explicit applied strain rate dependence has been introduced into the negative strain rate sensitivity of the flow stress.  The method should also be applicable in intermittent flow observed in worm like micellar systems \cite{Sood06}.

Some comments are in order on the scope of the model. Comparison with experiments is difficult due to paucity of experimental results, in particular since most parameters that in principal can affect the dynamics are kept constant. For instance,  our investigations show that most theoretical parameters such as the roller inertia, tape mass, visco-elastic parameters like $k_g(0)$, $k_g(\infty)$ and $T_a$ (connected to the complex compliance)  that affect the dynamics, are experimentally accessible parameters. However, conventional experiments are performed keeping these parameters fixed \cite{MB,BC97,CGVB04}, presumably as there has been no suggestion that these parameters would affect the peel dynamics. It would be interesting to test the prediction of the model by altering these parameters.  While changing roller inertia or tape mass is straight-forward, there is no reference material with respect which visco-elastic contribution can introduced. The best that can be done is to study the changes in the dynamics by using tapes manufactured with different adhesives.  Moreover, the available experimental results are  mostly on acoustic emission measured as a function of pull velocity keeping all other parameters fixed. This was dealt in our  earlier publications \cite{Jag08a,Jag08b}.   Finally, it should be stated that effects arising from thickness of the adhesive film are beyond the scope of the model.

\section*{Appendix}
\appendix
\renewcommand\thesection{A-\Alph{section}}
\setcounter{equation}{0}

Consider a system that organizes into a critical state under driving. Let the size of event  denoted by   $s$  occur in a  duration $T$. Then, both these quantities follow a power-law distribution defined by 

\begin{eqnarray}
P(s) \sim s^{-\alpha},\\
\label{eventA}
P(T) \sim T^{-\beta}.
\label{durationA}
\end{eqnarray}
The lifetime of an event $T$ is related to its size $s$ by
\begin{equation}
s \sim T^x.
\label{lifetimeA}
\end{equation}

Clearly, event sizes and their durations are not  independent and therefore all the three exponents are not independent. Indeed,   a proper statistical description requires that we use the joint  probability density $P(s,T) ds dT$ of having signals  with amplitudes between $s$ and $s+ds$ occurring in a duration with $T$ and $T+dT$ \cite{Vives}. Using $P(s,T)$, a scaling relation  between the three exponents has been derived for the case when the event sizes and durations exhibit a single scaling regime \cite{Vives}. Following Ref. \cite{Vives}, we  derive scaling relations valid for a two stage power law distribution.

Given the joint  probability density $P(s,T)$,  the two marginal probability densities are given by
\begin{eqnarray}
\nonumber 
P(s) &= &\int_{T_{min}}^{T_{max}} P(s,T)dT, \\
P(T)& =& \int_{s_{min}}^{s_{max}} P(s,T)ds,
\end{eqnarray}
where $T_{max}$, $T_{min}$, $s_{max}$, and $s_{min}$ are the upper and lower 
cutoffs for $T$ and $s$ imposed by the  particular experimental setup 
 within which $P(s,T)$ is normalized, i.e.,
\begin{equation}
\int_{T_{min}}^{T_{max}} \int_{s_{min}}^{s_{max}} P(s,T)dsdT = 1.
\end{equation}

For the current situation, we assume 
\begin{equation}
P(T) \sim  T^{-\beta} P_C(T) \sim T^{-\beta} \frac{1}{1+ A^2 T^2},
\label{duration_twoA}
\end{equation}
instead of Eq. (\ref{durationA}). Clearly the exponent  
\begin{eqnarray}
\beta &= &\beta_1  \, \, {\rm for \,\,} A^2T^2 << 1, 
\end{eqnarray}
corresponds to the first region of scaling, while for the second, we have 
\begin{equation}
\beta_2 = \beta_1 + 2,  \, \,  {\rm for \,\,} A^2T^2 >> 1.
\end{equation}
The above choice (Eq. \ref{duration_twoA}) is motivated by some general considerations. We first note that the distribution must be well behaved for large $T$. Second, the event sizes and their durations corresponding to the second scaling regime are likely to be uncorrelated, particularly in time. In the context of peeling, large acoustic emission bursts require large segments to be peeled almost simultaneously. Such events are likely to be well separated in time and  therefore such events are likely to act as independent events. Finally, the functional form  $P_C(T) \sim \frac{1}{1+ A^2 T^2}$ is the well known Cauchy distribution.  This specific choice  is motivated by the fact that the Cauchy distribution is one of the distributions that reproduces itself under addition of identically distributed independent random variables.   It is clear that this choice gives the exponent value  $\sim \beta$ for $T << 1/A$,  while for $ T>> 1/A$, the exponent is close to $2 + \beta$. There would be a cross-over region around $ T= T^* \sim 1/A$. 

With this, for a two stage scaling regime, we interpret Eq. (\ref{eventA}) to imply 
$\alpha = \alpha_1$ for the first scaling regime seen at small values of $s$ and $\alpha=\alpha_2$ for second scaling regime of large values of $s$. In contrast, we assume a single regime  relating event size $s$ with its duration $T$ with a scaling exponent $x$.  
 
Then, combining  Eqs. (\ref{duration_twoA},\ref{lifetimeA},\ref{eventA}) ( with the above interpretation), a  general scaling form for $P(s,T)$ can be written as,
\begin{equation}
P(s,T)=g(s/T^{x}) s^{-\theta} f(T),
\end{equation}
where $\theta$ is an exponent. We assume that the function $g(z)$ (with $z=s/T^x$) is a ``well-localized'' distribution function with a maximum around  $z_0$ and strongly decaying on either side of $z$.  

Note that the scaling variable $z=s/T^{x}$ gives a precise definition  for the exponent $x$. The functions $g$ and $f(T)$ can always be redefined so  that $P(s,T)=G(s/T^{x}) \phi(T)$. In this case, under the change
 $s \rightarrow z = s/T^{x}$ reads
\begin{equation}
P(T)=\phi(T) T^{x} \int_{\frac{ s_{min}}{T^{x}} }^{ \frac{s_{max}}{T^{x}} } G(z)dz.
\label{tz}
\end{equation}
On comparing this equation  with Eqn. (\ref{duration_two}), we get 
\begin{equation}
\phi(T) \sim \frac{T^{-\beta-x}}{1 + A^2T^2}.\\
\label{T} 
\end{equation}

Using Eq. (\ref{T}), we can  rewrite $P(s)$ as
\begin{equation}
P(s)= \int_{T_{min}}^{T_{max}} G(s/T^x) \frac{T^{-\beta-x}}{1 + A^2T^2}dT.
\end{equation}
Change of variables from $s$ to  $z= s/T^x$ leads to 
\begin{equation}
P(s)= \int_{(s/z_{max})^{1/x}}^{(s/z_{min})^{1/x}}G(z) \frac{s^{-(\beta+x-1)/x} z^{-(1+x)/x}}{1 + A^2(s/z)^{2/x}}dz.
\end{equation}
By limiting the range of integration to appropriate limits, the equation can be seen to have  two regions of scaling. The first one is for small $s$
\begin{equation}
P(s) \sim s^{-\alpha_1} \sim   s^{-(\beta +x -1)/x} , \, \, {\rm for } \,\, A^2(s/z)^{2/x} << 1,
\label{event_s}
\end{equation}
which gives 
\begin{equation}
x(1-\alpha_1) = 1-\beta_1.
\label{singlePL}
\end{equation}
This is the standard scaling relation when the distribution exhibits a single power law. For large $s$, we get
\begin{equation}
P(s) \sim s^{-\alpha_2} \sim  s^{-(\beta +x +1)/x}  ,\, \, {\rm for } \,\, A^2(s/z)^{2/x} >> 1.
\label{event_l}
\end{equation}
which gives 
\begin{equation}
x(\alpha_2 -1) = \beta + 1.
\label{twoPL}
\end{equation}
with $ \beta = \beta_1$ corresponding to the first scaling region.  It is important to note that the exponent corresponding to event size for the second scaling regime $\alpha_2$ is completely determined in terms of the $\beta_1$ of the first stage and $x$. ( Note also $\beta_2 = \beta_1 +2$.) Further, we stress that the above derivation makes no reference to slow driving at all. Indeed, in the case of the PLC effect,  the power laws are seen at high drive rates much like in hydrodynamics \cite{Anan99,Bharepl,Anan04}. In the present case, power law distributions  are seen at low as well as high drive rates.

\centerline{
{\bf ACKNOWLEDGMENTS}}
GA would like to acknowledge the grant of Raja Ramanna Fellowship and also support from BRNS Grant No. $2007/36/62$-$BRNS/2564$ .


\begin{thebibliography}{99}
\bibitem{MB} D. Maugis and M. Barquins in {\it Adhesion 12}, edited by K. W. Allen (Elsevier, London, 1988), p. 205.

\bibitem{BC97} M. Barquins and M. Ciccotti, Int. J. Adhes. Adhes. {\bf 17}, 65 (1997).

\bibitem{CGVB04} M. Ciccotti, B. Giorgini, D. Villet, and M. Barquins, Int. J. Adhes. Adhes. {\bf 24}, 143 (2004).

\bibitem{McEwan} A. D. McEwan, Rheologica Acta {\bf 5}, 205 (1966).

\bibitem{Urahama} Y. Urahama, J. Adhes. {\bf 31}, 47 (1989).

\bibitem{Dickinson} L. Scudiero, I. T. Dickinson, L. C. Jensen and S. C. Langford,  J. Adhes. Sci. Technol. {\bf 9}, 27 (1995).

\bibitem{Rumi06} Rumi De and G. Anantahakrishna, Phys. Rev. Lett. {\bf 97}, 165503 (2006).

\bibitem{Jag08a} Jagadish Kumar, M. Ciccotti, and G. Ananthakrishna, Phys. Rev. E {\bf 77}, 045202(R) (2008).

\bibitem{Jag08b} Jagadish Kumar, Rumi De, and G. Ananthakrishna, Phys. Rev. E {\bf 78}, 066119 (2008).

\bibitem{Heslot94} F. Heslot {\it et al}.,  Phys. Rev. E {\bf 49}, 4973 (1994).

\bibitem{Persson} B. N. J. Persson, {\it Sliding Friction: Physical Principles and Applications}, 2nd ed. (Springer, Heidelberg, 2000).

\bibitem{PLC} A. Portevin and F. Le Chatelier, C. R. Acad. Sci. Paris {\bf 176},
507 (1923); F. Le Chatelier, Re. de Metallurgie {\bf 6}, 914 (1909).

\bibitem{GA07} G. Ananthakrishna, Phys. Rep. {\bf 440}, 113 (2007).

\bibitem{Sood06} Rajesh Ganapathy and A. K. Sood,  Phys. Rev. Lett. {\bf 96}, 108301 (2006); N. A. Spenley, M. E. Cates and T. C. B. Mcleish, Phys. Rev. Lett. {\bf 71}, 939 (1993).

\bibitem{Anan04} G. Ananthakrishna and M. S. Bharathi, Phys. Rev. E {\bf 70}, 26111 (2004).

\bibitem{Kubin86} L. P. Kubin and Y. Estrin, J. de Physique {\bf 47}, 497 (1986).

\bibitem{Kae64} D. H. Kaelble, J. Colloid. Sci. {\bf 19}, 413 (1964).

\bibitem{GP69} A. N. Gent, R. P. Petrich, Proc. Roy. Soc. {\bf A 310}, 433 (1969).

\bibitem{TIY04} M. Takiguchi, S. Izumi and F. Yoshida, Proc. Instn. Mech. Engrs., part C: J. Mechanical Engineering Science, {\bf 218}, 623 (2004).

\bibitem{BKC} See {\it Modelling Critical and Catastrophic Phenomena in Geosciences: a Statistical Physics Approach}, edited by P. Bhattacharyya and B. K. Chakrabarti, Lect. Notes Phys. Vol. 705 (Springer, Berlin, 2006).

\bibitem{Anan06} G. Ananthakrishna and Rumi De in Ref. \cite{BKC}.

\bibitem{YZZ04} S. Yand, Y-W. Zhang and K. Zeng, J. Appl. Phys. {\bf 95}, 3655 (2004).

\bibitem{Rumi04} Rumi De, Anil Maybhate and G. Ananthakrishna, Phys. Rev. E {\bf 70}, 046223 (2004).

\bibitem{Kubin} M. A. Lebyodkin, Y. Brechet, Y. Estrin and L. P. Kubin, Phys. Rev.Lett. {\bf 74}, 4758 (1995).

\bibitem{Rumiepl} Rumi De and G. Ananthakrishna, Europhys. Lett. {\bf 66}, 715 (2004).

\bibitem{vives} E. Vives, J. Ort\'in, L. Ma\~nosa, I. R\'afols, R. P\'erez-Magran\'e, and A. Planes Phys. Rev. Lett. {\bf 72 }, 1694 (1994).

\bibitem{Rajeev01} R. Ahluwalia and G. Ananthakrishna, Phys. Rev. Lett. {\bf 86}, 4076 (2001).

\bibitem{Kala03} S. Sreekala and G. Ananthakrishna, Phys. Rev. Lett. {\bf 90}, 135501 (2003).

\bibitem{Rumi05} Rumi De and G. Ananthakrishna, Phys. Rev. E {\bf 71}, 055201(R) (2005).

\bibitem{ECG} E. C. G. Sudarshan and N. Mukunda, {\it Classical Dynamics: A Modern Perspective} (Wiley, New York, 1974).

\bibitem{Land}L. D. Landau and E. M. Lifschitz, {\it Theory of Elasticity} (Pergamon, Oxford, 1986).

\bibitem{Bohr}T. Bohr, M.H. Jensen, G. Paladin and A. Vulpiani, {\it Dynamical Systems Approach to Tubulence}, (Cambridge,United Kingdom, 1997 ).

\bibitem{Yamazaki} Y. Yamazaki and A. Toda, Physica D {\bf 214}, 120 (2006).

\bibitem{Vives} I. R\'afols and E. Vives, Phys. Rev. B {\bf 52}, 12651 (1995). 

\bibitem{Anan99} G. Ananthakrishna, S. J. Noronha, C. Fressengeas, and L. P. Kubin, Phys. Rev. E {\bf 60}, 545.pdf5 (1999).

\bibitem{Bharepl} M. S. Bharathi and G. Ananthakrishna, Euro. Phys. lett. {\bf 60}, 234(2002).
\end{thebibliography}
\end{document}